\documentclass[12pt]{article}
\usepackage{epsfig,cite}
\include{epsf}
\newcount\timecount
\newcount\hours \newcount\minutes  \newcount\temp \newcount\pmhours
\hours = \time
\divide\hours by 60
\temp = \hours
\multiply\temp by 60
\minutes = \time
\advance\minutes by -\temp
\def\hour{\the\hours}
\def\minute{\ifnum\minutes<10 0\the\minutes
            \else\the\minutes\fi}
\def\clock{
\ifnum\hours=0 12:\minute\ AM
\else\ifnum\hours<12 \hour:\minute\ AM
      \else\ifnum\hours=12 12:\minute\ PM
            \else\ifnum\hours>12
                 \pmhours=\hours
                 \advance\pmhours by -12
                 \the\pmhours:\minute\ PM
                 \fi
            \fi
      \fi
\fi
}

\def\monthname{\relax\ifcase\month 0/\or January\or February\or
   March\or April\or May\or June\or July\or August\or September\or
   October\or November\or December\else\number\month/\fi}

\def\bold#1{\setbox0=\hbox{$#1$}%
     \kern-.025em\copy0\kern-\wd0
     \kern.05em\copy0\kern-\wd0
     \kern-.025em\raise.0433em\box0 }

\def\beq{\begin{equation}}
\def\eeq{\end{equation}}

\def\ga{\mathrel{\raise.3ex\hbox{$>$\kern-.75em\lower1ex\hbox{$\sim$}}}}
\def\la{\mathrel{\raise.3ex\hbox{$<$\kern-.75em\lower1ex\hbox{$\sim$}}}}
\def\gev{{\rm \, Ge\kern-0.125em V}}
\def\tev{{\rm \, Te\kern-0.125em V}}
\def\gyr{{\rm \, G\kern-0.125em yr}}

\def\gappeq{\mathrel{\rlap {\raise.5ex\hbox{$>$}}
{\lower.5ex\hbox{$\sim$}}}}
\def\lappeq{\mathrel{\rlap{\raise.5ex\hbox{$<$}}
{\lower.5ex\hbox{$\sim$}}}}
\def\Toprel#1\over#2{\mathrel{\mathop{#2}\limits^{#1}}}

\def\m12{m_{1\!/2}}

\def\PL{{Phys.~Lett.} }
\def\PR{{Phys.~Rev.} }

\def\bea{\begin{eqnarray}}
\def\eea{\end{eqnarray}}

\def\Yi{\eta^{\ast}_{11} \left( \frac{y_{i}}{2} g' Z_{\chi 1} +
        g T_{3i} Z_{\chi 2} \right) + \eta^{\ast}_{12}
        \frac{g m_{q_{i}} Z_{\chi 5-i}}{2 m_{W} B_{i}}}

\def\Xii{\eta^{\ast}_{11}
        \frac{g m_{q_{i}}Z_{\chi 5-i}^{\ast}}{2 m_{W} B_{i}} -
        \eta_{12}^{\ast} e_{i} g' Z_{\chi 1}^{\ast}}

\def\Wi{\eta_{21}^{\ast}
        \frac{g m_{q_{i}}Z_{\chi 5-i}^{\ast}}{2 m_{W} B_{i}} -
        \eta_{22}^{\ast} e_{i} g' Z_{\chi 1}^{\ast}}

\def\Vi{\eta_{22}^{\ast} \frac{g m_{q_{i}} Z_{\chi 5-i}}{2 m_{W} B_{i}}
        + \eta_{21}^{\ast}\left( \frac{y_{i}}{2} g' Z_{\chi 1}
        + g T_{3i} Z_{\chi 2} \right)}

\def\zthree{\delta_{1i} [g Z_{\chi 2} - g' Z_{\chi 1}]}

\def\zfour{\delta_{2i} [g Z_{\chi 2} - g' Z_{\chi 1}]}

\begin{document}
\begin{titlepage}
\pagestyle{empty}
\baselineskip=21pt
\rightline{\tt hep-ph/0502001}
\rightline{CERN-PH-TH/2005-002}
\rightline{UMN--TH--2342/05}
\rightline{FTPI--MINN--05/01}
\vskip 0.2in
\begin{center}
{\large {\bf Update on the Direct Detection of Supersymmetric Dark 
Matter}}
\end{center}
\begin{center}
\vskip 0.2in
{\bf John~Ellis}$^1$, {\bf Keith~A.~Olive}$^{2}$, {\bf Yudi~Santoso}$^{3}$ 
and {\bf Vassilis~C.~Spanos}$^{2}$
\vskip 0.1in
{\it
$^1${TH Division, CERN, Geneva, Switzerland}\\
$^2${William I. Fine Theoretical Physics Institute, \\
University of Minnesota, Minneapolis, MN 55455, USA}\\
$^3${Department of Physics, University of Guelph, Guelph, ON N1G 2W1, Canada}\\
{and Perimeter Institute of Theoretical Physics,  Waterloo, ON N2L 2Y5, Canada}
}\\
\vskip 0.2in
{\bf Abstract}
\end{center}
\baselineskip=18pt \noindent

We compare updated predictions for the elastic scattering of
supersymmetric neutralino dark matter with the improved
experimental upper limit recently published by CDMS~II. We take into
account the possibility that the $\pi$-nucleon $\Sigma$ term may be
somewhat larger than was previously considered plausible, as may be
supported by the masses of exotic baryons reported recently. We also
incorporate the new central value of $m_t$, which affects indirectly
constraints on the supersymmetric parameter space, for example via
calculations of the relic density. Even if a large value of $\Sigma$ is
assumed, the CDMS~II data currently exclude only small parts of the parameter space
in the constrained MSSM (CMSSM) with universal soft supersymmetry-breaking
Higgs, squark and slepton masses. None of the previously-proposed CMSSM
benchmark scenarios is excluded for any value of $\Sigma$, and the CDMS~II
data do not impinge on the domains of the CMSSM parameter space favoured
at the 90\% confidence level in a recent likelihood analysis. However,
some models with non-universal Higgs, squark and slepton masses and
neutralino masses $\lappeq 700$~GeV are excluded by the CDMS~II data.

\vfill
\leftline{CERN-PH-TH/2005-002}
\leftline{January 2005}
\end{titlepage}
\baselineskip=18pt

\section{Introduction}

The lightest supersymmetric particle (LSP) is stable in models in which
$R$ parity is conserved, in which case it is a suitable candidate for the
cold dark matter required by astrophysical and cosmological observations \cite{EHNOS}.
One of the generic possibilities is that the LSP is the lightest
neutralino $\chi$, in which case the detection of dark matter appears
feasible. The direct detection of supersymmetric dark matter via
scattering on nuclei in deep-underground, low-background experiments has
been discussed many times \cite{etal} - \cite{bot2}. 

There are, however, three reasons why a re-evaluation of the prospects for
such experiments is now timely. The first is the motivation provided by
the upper limit on the dark-matter scattering cross section provided by
the CDMS~II experiment \cite{CDMS2}, which is substantially more stringent than
previous experiments \cite{prev}. The CDMS~II result appears, in particular, to
conflict with the dark-matter scattering interpretation of the results of
the previous DAMA experiment \cite{dama}. A second reason is evolution in Standard
Model inputs into the calculation of the scattering matrix elements.
Recent particle-physics experiments tend to favour a value of the
pion-nucleon sigma term $\Sigma$ that is somewhat higher than earlier
experiments, favouring a larger theoretical estimate for the
spin-independent part of the dark-matter scattering cross section \cite{highsnp}.
Interestingly, a larger value of $\Sigma$ is also favoured independently
by hints from the spectroscopy of pentaquark baryons, if they exist \cite{exotic}. We
also include the effect of the new preferred value of $m_t$ \cite{mtop} on the
supersymmetric parameter space and on relic-density calculations. Finally,
there has been some progress recently in understanding which parts of
parameter space are favoured in certain versions of the minimal
supersymmetric extension of the Standard Model (MSSM). In particular, if
the input soft supersymmetry-breaking parameters are constrained to be
universal (CMSSM), the data on $m_W, \ \sin^2 \theta_W$ and $g_\mu - 2$ all
favour independently a relatively low mass for the lightest 
neutralino~\cite{EHOW3}, favouring in turn a relatively large
dark-matter scattering cross section.

It is the purpose of this paper, in light of these developments, to
re-evaluate the prospects for discovering dark-matter scattering in
forthcoming experiments. We include in our analysis not only models in
which neutralinos are the dominant source of cold dark matter, but also
those in which neutralinos provide only some fraction $f_\chi < 1$. In the
latter case, we assume that neutralinos constitute the same fraction
$f_\chi < 1$ of the galactic halo. For comparison with experiments
searching for dark-matter scattering, which usually assume that all the
halos are composed of neutralinos, we rescale the effective scattering
cross section by the same factor $f_\chi < 1$.

We find that, even with the larger value of $\Sigma$, only very small
parts of the CMSSM parameter space are excluded by the current CDMS~II
result. Specifically, none of the benchmark scenarios proposed
recently~\cite{Bench2} is excluded, and neither is any of the 90\%
confidence-level region favoured in a recent likelihood analysis of the
CMSSM~\cite{EHOW3}. On the other hand, if one relaxes universality for the
squark slepton and Higgs masses, so as to consider the most general
low-energy effective supersymmetric theory (LEEST), some models with
$m_\chi \lappeq 700$~GeV are excluded for large $\Sigma$. We reach a
similar conclusion even if the squark and slepton masses are assumed to be
equal, and we allow only non-universal Higgs masses (NUHM). Indeed, as we
discuss, the dominant mechanism leading to a large cross section is the
reduction in the magnitude of the Higgs superpotential mixing parameter
$\mu$ and the pseudoscalar Higgs mass $m_A$ allowed by the relaxed 
electroweak vacuum conditions in the NUHM.

\section{Spin-Independent $\chi$-Nucleon Scattering Matrix Elements}

\subsection{Model-Dependent Supersymmetric Operator Coefficients}

We assume that the neutralino LSP $\chi$ is the
lightest eigenstate of the mixed Bino ${\tilde B}$, Wino $\tilde 
W$
and Higgsino ${\tilde H}_{1,2}$ system, whose mass matrix $N$ is
diagonalized by a matrix $Z$: $diag(m_{\chi_1,..,4}) = Z^* N Z^{-1}$, with
\begin{equation}
\chi = Z_{\chi 1}\tilde{B} + Z_{\chi 2}\tilde{W} +
Z_{\chi 3}\tilde{H_{1}} + Z_{\chi 4}\tilde{H_{2}}.
\label{id}
\end{equation}
We neglect the possibility of CP violation, and assume
universality at the supersymmetric GUT scale for the
$U(1)$ and $SU(2)$ gaugino masses: $M_{1,2} = m_{1/2}$, so that
$M_1 = \frac{5}{3}\tan^2{\theta_{W}}M_{2}$ at the electroweak scale.

The following low-energy effective
four-fermion
Lagrangian describes spin-independent elastic $\chi$-nucleon   
scattering:
\begin{equation}
{\cal L} \, = \, \alpha_{3i} \bar{\chi} \chi \bar{q_{i}} q_{i},
\label{lagr}
\end{equation}
which is to be summed over the quark flavours $q$, and the
subscript $i$ labels up-type quarks ($i=1$) and down-type quarks
($i=2$). The model-dependent coefficients $\alpha_{3i}$ are given by
\begin{eqnarray}
\alpha_{3i} & = & - \frac{1}{2(m^{2}_{1i} - m^{2}_{\chi})} Re \left[
\left( X_{i} \right) \left( Y_{i} \right)^{\ast} \right]
- \frac{1}{2(m^{2}_{2i} - m^{2}_{\chi})} Re \left[
\left( W_{i} \right) \left( V_{i} \right)^{\ast} \right] \nonumber \\
& & \mbox{} - \frac{g m_{qi}}{4 m_{W} B_{i}} \left[ Re \left(  
\zthree \right) D_{i} C_{i} \left( - \frac{1}{m^{2}_{H_{1}}} +   
\frac{1}{m^{2}_{H_{2}}} \right) \right. \nonumber \\
& & \mbox{} +  Re \left. \left( \zfour \right) \left(
\frac{D_{i}^{2}}{m^{2}_{H_{2}}}+ \frac{C_{i}^{2}}{m^{2}_{H_{1}}}
\right) \right] ,
\label{alpha3}
\end{eqnarray}
where
\begin{eqnarray}
X_{i}& \equiv& \Xii , \nonumber \\
Y_{i}& \equiv& \Yi , \nonumber \\
W_{i}& \equiv& \Wi , \nonumber \\
V_{i}& \equiv& \Vi ,
\label{xywz}
\end{eqnarray}
with $y_i, T_{3i}$ denoting hypercharge and isospin, and
\beq
\delta_{1i} = Z_{\chi 3} (Z_{\chi 4}), \qquad \delta_{2i} = Z_{\chi 4},
(-Z_{\chi 3})
\eeq
\beq
B_{i} = \sin{\beta} (\cos{\beta}) , \qquad C_{i} = \sin{\alpha} (\cos{\alpha}), 
\qquad D_{i} = \cos{\alpha} ( - \sin{\alpha}),
\label{moredefs}
\eeq
for up (down) type quarks. 
We denote by $m_{H_2} < m_{H_1}$
the two scalar Higgs masses, and $ \alpha $ denotes the Higgs mixing
angle. Finally, we note that the factors $\eta_{ij}$ arise from the 
diagonalization of the squark mass matrices: $diag(m^2_1, m^2_2) \equiv
\eta M^2 \eta^{-1}$, which can be parameterized
for each flavour $f$ by an angle $ \theta_{f} $ and phase $\gamma_f$:
\begin{equation}
\left( \begin{array}{cc} 
\cos{\theta_{f}} & \sin{\theta_{f}} e^{i \gamma_{f}} \nonumber \\
-\sin{\theta_{f}} e^{-i \gamma_{f}} & \cos{\theta_{f}}
\end{array} \right)
\hspace{0.5cm}
 \equiv
\hspace{0.5cm}
 \left( \begin{array}{cc}
\eta_{11} & \eta_{12} \nonumber \\
\eta_{21} & \eta_{22}
\end{array} \right).
\label{defineeta}
\end{equation}
In the models we study below, the squark flavours are diagonalized in the 
same basis as the quarks.

\subsection{Hadronic Matrix Elements}
 
The scalar part of the
cross section can be written as
\begin{equation}
\sigma_{3} = \frac{4 m_{r}^{2}}{\pi} \left[ Z f_{p} + (A-Z) f_{n}
\right]^{2} ,
\label{si}
\end{equation}
where $m_r$ is the reduced LSP mass,
\begin{equation}
\frac{f_{p}}{m_{p}} = \sum_{q=u,d,s} f_{Tq}^{(p)}
\frac{\alpha_{3q}}{m_{q}} +
\frac{2}{27} f_{TG}^{(p)} \sum_{c,b,t} \frac{\alpha_{3q}}{m_q},
\label{f}
\end{equation}
the parameters $f_{Tq}^{(p)}$  are defined by
\begin{equation}
m_p f_{Tq}^{(p)} \equiv \langle p | m_{q} \bar{q} q | p \rangle
\equiv m_q B_q ,
\label{defbq}
\eeq
$f_{TG}^{(p)} = 1 - \sum_{q=u,d,s} f_{Tq}^{(p)} $~\cite{SVZ},
and $f_{n}$ has a similar expression.  

We take the ratios of the quark masses from~\cite{leut}:
\beq
{m_u \over m_d} = 0.553 \pm 0.043 , \qquad
{m_s \over m_d} = 18.9 \pm 0.8,
\eeq
and following~\cite{Cheng}, we have:
\beq
z \equiv {B_u - B_s \over B_d - B_s} = 1.49.
\label{chengvalue}
\eeq
Defining
\beq
y \equiv {2 B_s \over B_d + B_u},
\label{defy}
\eeq
we then have
\beq
{B_d \over B_u} = {2 + (z - 1) y \over 2 z - (z - 1) y}.
\label{bdoverbu}
\eeq
The coefficients $f_{T_q}$ are then easily obtained;
\beq
f_{T_u} = {m_u B_u \over m_p} = {2 \Sigma \over m_p (1+ {m_d \over m_u}) 
(1 + {B_d \over B_u})} ,
\eeq
\beq
f_{T_d} = {m_d B_d \over m_p} = {2 \Sigma \over m_p (1+ {m_u \over m_d}) 
(1 + {B_u \over B_d})} ,
\eeq
\beq
f_{T_s} = {m_s B_s \over m_p} = {2 ({m_s \over m_d}) \Sigma \, y \over m_p 
(1+ {m_u \over m_d}) } .
\eeq
The final task is to determine the quantity $y$ characterizing the density 
of $\bar{s}s$ in the nucleon.

This may be determined from the $\pi$-nucleon $\Sigma$ term, which is 
given by
\beq
\sigma_{\pi N} \equiv \Sigma = {1 \over 2} (m_u + m_d) (B_u + B_d) .
\eeq
We are motivated to reconsider the value of $y$ in light of recent 
re-evaluations of 
the $\pi$-nucleon sigma term $\Sigma$, which is related to the strange 
scalar density in the nucleon by
\begin{equation}
y = 1 - \sigma_0/\Sigma,
\end{equation}
where $\sigma_0$ is the change in the nucleon mass due to the 
non-zero $u, d$ quark masses, which is estimated on the basis of octet 
baryon mass differences to be $\sigma_0 = 36 
\pm 7$ MeV~\cite{oldsnp}. In our previous work~\cite{EFlO1,EFlOSo}, we assumed a 
relatively 
conservative value $\Sigma = 45$~MeV, which was already somewhat larger 
than naive quark model estimates, and corresponded to $y \simeq 0.2$. 
However, recent determinations of the $\pi$-nucleon $\Sigma$ term have
found the following values at the Cheng-Dashen point $t = + 2 m_\pi^2$ \cite{highsnp}:
\begin{equation}
\Sigma_{CD} \, = \, (88 \pm 15, \, 71 \pm 9, \, 79 \pm 7, \, 85 \pm 
5)~{\rm MeV}.
\label{sigmaCD}
\end{equation}
These should be corrected by an amount $- \Delta_R - \Delta_\sigma \simeq 
- 15$~MeV to obtain $\Sigma$. Assuming for definiteness the value 
$\Sigma_{CD} = 79 \pm 7$~MeV, we may estimate
\begin{equation}
\Sigma \, = \, (64 \pm 8) \, {\rm MeV}.
\label{sigma}
\end{equation}
Such a relatively large value of 
$\Sigma$ has recently received support from an unexpected 
quarter, namely 
the apparent observation of exotic baryons $\Theta^+, \Xi^{--}$ in an 
antidecuplet of flavour SU(3) \cite{exotic}. The existence of such states has been a 
long-standing prediction of chiral-soliton models, but the details of 
their spectroscopy depend, in particular, on the value of $\Sigma$:
\begin{eqnarray}
\frac{m_s}{m} \, \Sigma &  = &
\underbrace{3(4M_{\Sigma} - 3M_{\Lambda} - M_N)} +
\underbrace{4(M_{\Omega} - M_{\Delta})} \nonumber \\
& & \qquad \qquad octet \qquad \qquad \qquad  decuplet \nonumber \\
&  & -\underbrace{4(M_{\Xi^{--}} - M_{\Theta^+})} \nonumber \\
 & &  \quad \quad antidecuplet
\label{anti10spectrum}
\end{eqnarray}
in the chiral-soliton model.
Inserting the nominal values $M_{\Theta^+} = 1540$~MeV and $M_{\Xi^{--}} = 
1862$~MeV, we find $\Sigma = 72$~MeV, corresponding to $y \simeq 0.5$. 
This determination should be taken with a couple of grains of salt, since 
it is unclear whether either the $\Theta^+$ or the $\Xi^{--}$ exist. 
However, since this value is consistent with the more direct estimate
(\ref{sigma}), we adopt $\Sigma = 64$~MeV and 45~MeV as alternative 
nominal values, corresponding to $y \simeq 0.45$ and 0.2, respectively,
which we use later to discuss the implications of varying $\Sigma$.

\section{Exploration of the CMSSM}

We begin by considering the constrained version of the MSSM
(CMSSM)~\cite{cmssmnew,eoss,cmssmmap}. This class of models is fully
described by four parameters and a sign: a unified gaugino mass, $m_{1/2}$, a
unified scalar mass, $m_0$, a unified trilinear mass term, $A_0$, and the
ratio of the Higgs v.e.v.'s, $\tan \beta$. In addition, the sign of the
$\mu$ parameter must also be specified. The phenomenology of these models
has been well studied. The parameters of models with an acceptable
cosmological relic density generally fall into one of the following
regions: the coannihilation region~\footnote{Note that, because the relic
density has now been determined with high accuracy by cosmological
observations~\cite{wmap}, and accelerator limits disfavour low $m_{1/2}$,
we no longer distinguish a bulk region at low
$m_{1/2}$ and $m_0$ from the coannihilation region.}, where the mass of
the neutralino is nearly degenerate with the mass of the stau; the
rapid-annihilation funnel, where the mass of the neutralino is close to
one-half the mass of the pseudoscalar Higgs boson $A$; and the
focus-point region, which is found at extremely high values of $m_0$, and
is at the edge of the parameter space which allows for radiative
electroweak symmetry breaking. We start with an examination of some
specific benchmark parameter choices \cite{Bench2,EFFMO} that populate
these allowed regions. 

\subsection{Benchmark Scenarios}

Fig.~\ref{fig:sigma} shows the effect of the value of $\Sigma$ on the
magnitude of the spin-independent elastic $\chi$-proton scattering cross
section in the specific cases of the CMSSM benchmark scenarios discussed
in~\cite{Bench2}.  Points A, B, C, D, G, H, I, J, and L are in the
coannihilation region, points K and M are in the rapid-annihilation
funnels, and points E and F are in the focus-point region. Point M is not
shown as its cross section falls below the scale of the
plot~\footnote{These benchmark points were formulated assuming $m_t
=175$~GeV. The small shifts required if one uses the new central value
$m_t = 178$~GeV do not impact significantly the cross sections calculated
in Fig.~\ref{fig:sigma}.}. It is clear that the value of $\Sigma$ has
quite significant impact in all the scenarios, as indicated by the 
behaviours of the
different lines. There is a general trend for the cross section to
increase approximately quadratically with $\Sigma$. This would be exact if
the $<p|{\bar u} u|p>$ and $<p|{\bar d} d|p>$ contributions were
negligible compared with the $<p|{\bar s} s|p>$ contribution. However,
Fig.~\ref{fig:sigma} shows that the increasing trend is not exactly
universal, reflecting the different relative weights of the various
$<p|{\bar q} q|p>$ contributions in the different benchmark scenarios.  
These depend on $\tan \beta$ and the sign of the Higgs-mixing parameter
$\mu$, as can be seen from the formulae in the previous Section.

\begin{figure}[h]
\includegraphics[height=2.8in]{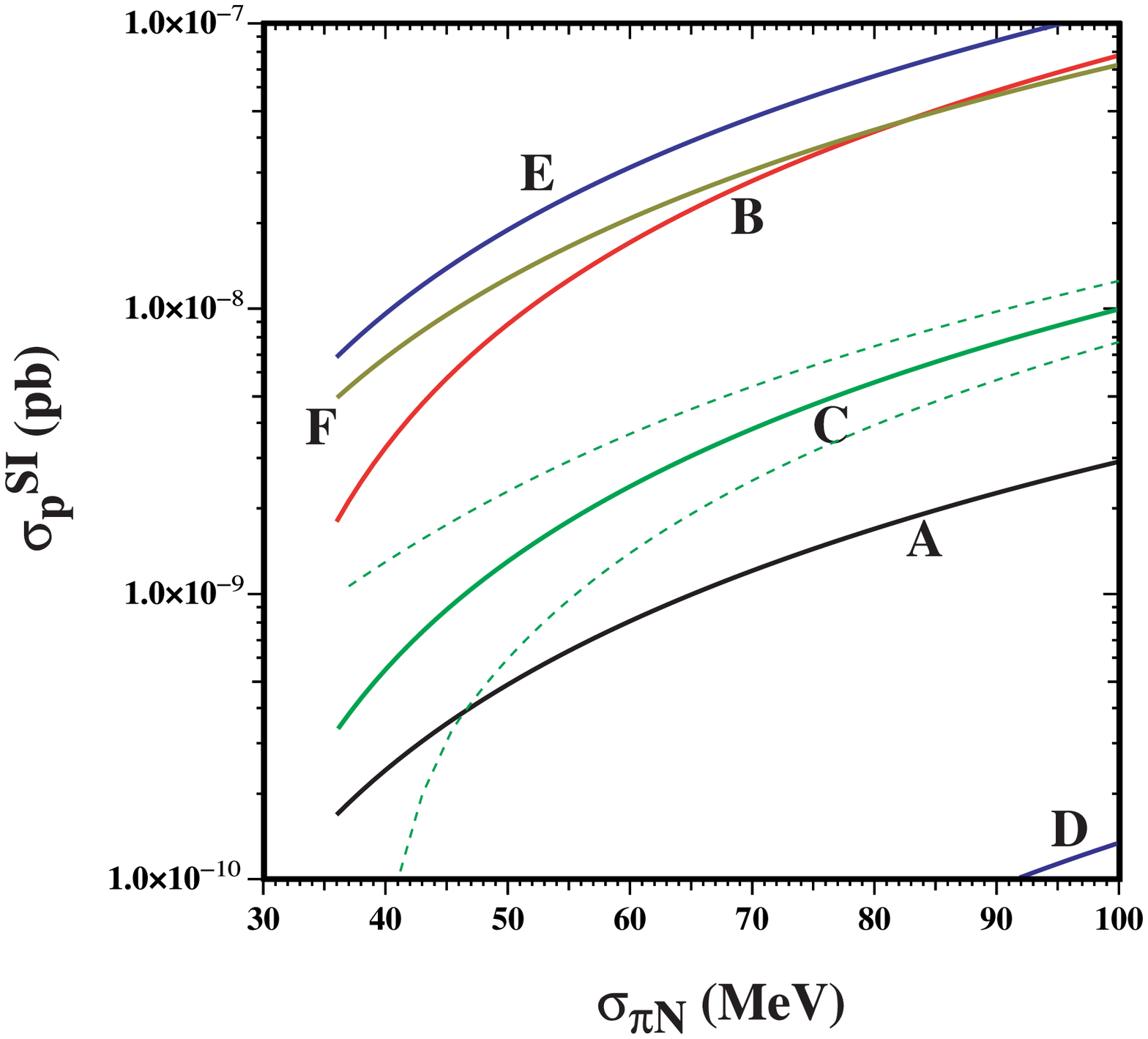}
\includegraphics[height=2.8in]{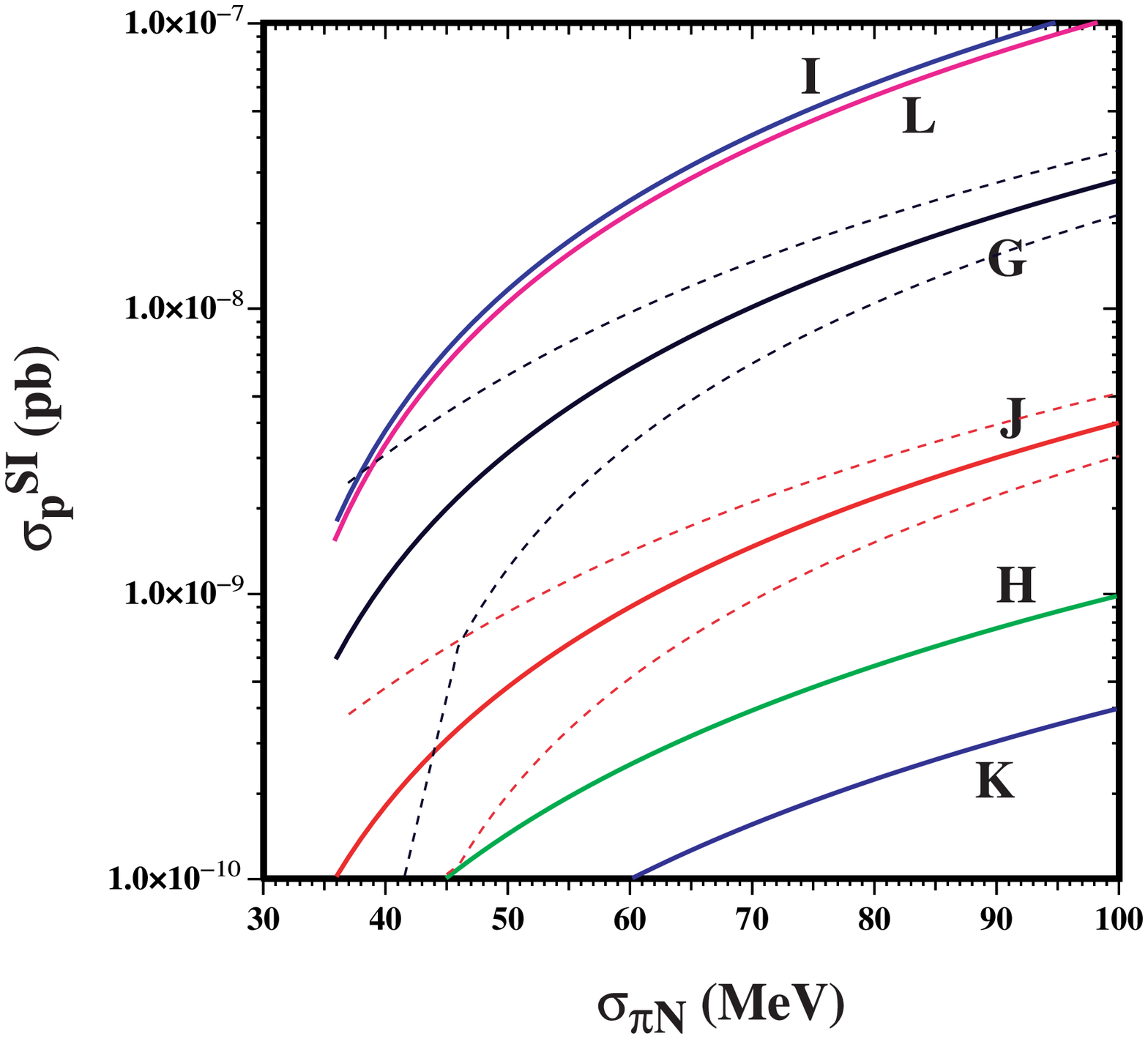}
\caption{\label{fig:sigma}
{\it The dependences on the $\pi$-nucleon $\Sigma$ term of the elastic 
cross sections of the benchmark points~\protect\cite{Bench2}. The dashed
lines indicate the sensitivities to $\sigma_0$ in the cases of benchmark 
scenarios C, G and J. The predicted cross sections are smaller than 
the CDMS~II upper 
limits~\protect\cite{CDMS2} for the the models considered, for all $\Sigma$ 
values shown.}} 
\end{figure}

We have plotted in Fig.~\ref{fig:sigma} values of the cross section
corresponding to $\Sigma \ge 36$ MeV, i.e., consistent with assuming $y
\ge 0$. The dashed curves in Fig.~\ref{fig:sigma} around benchmark points
C, G, and J display the effect of the uncertainty in $\sigma_0$ as well as
the mass ratios which enter into the determination of the $f_{T_q}$ and
ultimately the elastic cross section.  We see that this uncertainty is not
negligible, although that associated with $\Sigma$ is clearly more
important.

We see that, in all scenarios and for all plausible values of $\Sigma$,
the estimated cross section lies considerably below the current upper
limits of CDMS~II~\cite{CDMS2}, which can at best exclude models with
cross sections larger than $3 \times 10^{-7}$ pb when $m_\chi = 60$ GeV.
If future experiments achieve a sensitivity of $10^{-8}$ pb, one can
plainly see that several of the benchmark scenarios will be probed,
particularly if $\Sigma$ is large.

It is clear from the above discussion that better understanding of the
non-perturbative hadronic matrix elements $\Sigma$ and $\sigma_0$ will be
needed before the spin-independent elastic-scattering cross section can be
predicted accurately in any specific supersymmetric model. This means, in
particular, that {\it unless these hadronic matrix elements can be 
determined more accurately}, it will be difficult to convert any LHC or LC
measurements of MSSM parameters into accurate predictions for
elastic-scattering rates, even if they do suffice to calculate accurately
the relic LSP density. The experimental determination of $\Sigma$ is 
notoriously uncertain: perhaps the time is ripe for another lattice QCD 
approach?

The benchmark scenarios discussed above were formulated within the CMSSM,
and our next step is to explore the CMSSM more generally, to see 
whether larger cross sections are possible in regions of its 
parameter space.

\subsection{General Analysis of CMSSM Models compatible with WMAP}

As is well known, for any given value of $\tan \beta$, $A_0$ and $m_t$,
the CMSSM parameter space consists of narrow strips in the $(m_{1/2},
m_0)$ plane, where the relic density falls within the range allowed for
cold dark matter by WMAP and other experiments. In the following, we no
longer consider results in the focus-point region: this now appears at
very large $m_0$ if one adopts the new central value $m_t = 
178$~GeV~\footnote{We use $m_b (m_b)_{\overline{MS}}$ = 4.25~GeV 
throughout.}, as
we do henceforth. At low values of $m_{1/2}$, the length of the strip is
in turn restricted by experimental constraints such as $m_h$,
$m_{\chi^\pm}$ and $b \to s \gamma$, whereas at high values of $m_{1/2}$
the strips are truncated by the relic density.  We display in
Fig.~\ref{fig:m0m12} the $(m_{1/2}, m_0)$ planes for $\tan \beta = 10$ and
(a) $\mu < 0$, (b) $\mu > 0$, (c) $\tan \beta = 40, \ \mu < 0$ and (d) $\tan
\beta = 57, \ \mu > 0$. The latter choices of $\tan \beta$ are close to the
maximal values we now find for the corresponding signs of $\mu$. These
have increased with the new best-fit value $m_t = 178$~GeV and recent
improvements in our spectrum evaluation code~\footnote{The most recent 
improvements
include implementation of the full set of two-loop renormalization-group 
equations.}. The rapid-annihilation
funnels visible in panels (c, d) are located at values of $m_{1/2}$ that
are similar to what we would have found previously for $\tan \beta = 35,
\ \mu < 0$ and $\tan \beta = 50, \ \mu > 0$.

\begin{figure}
\begin{center}
\mbox{\epsfig{file=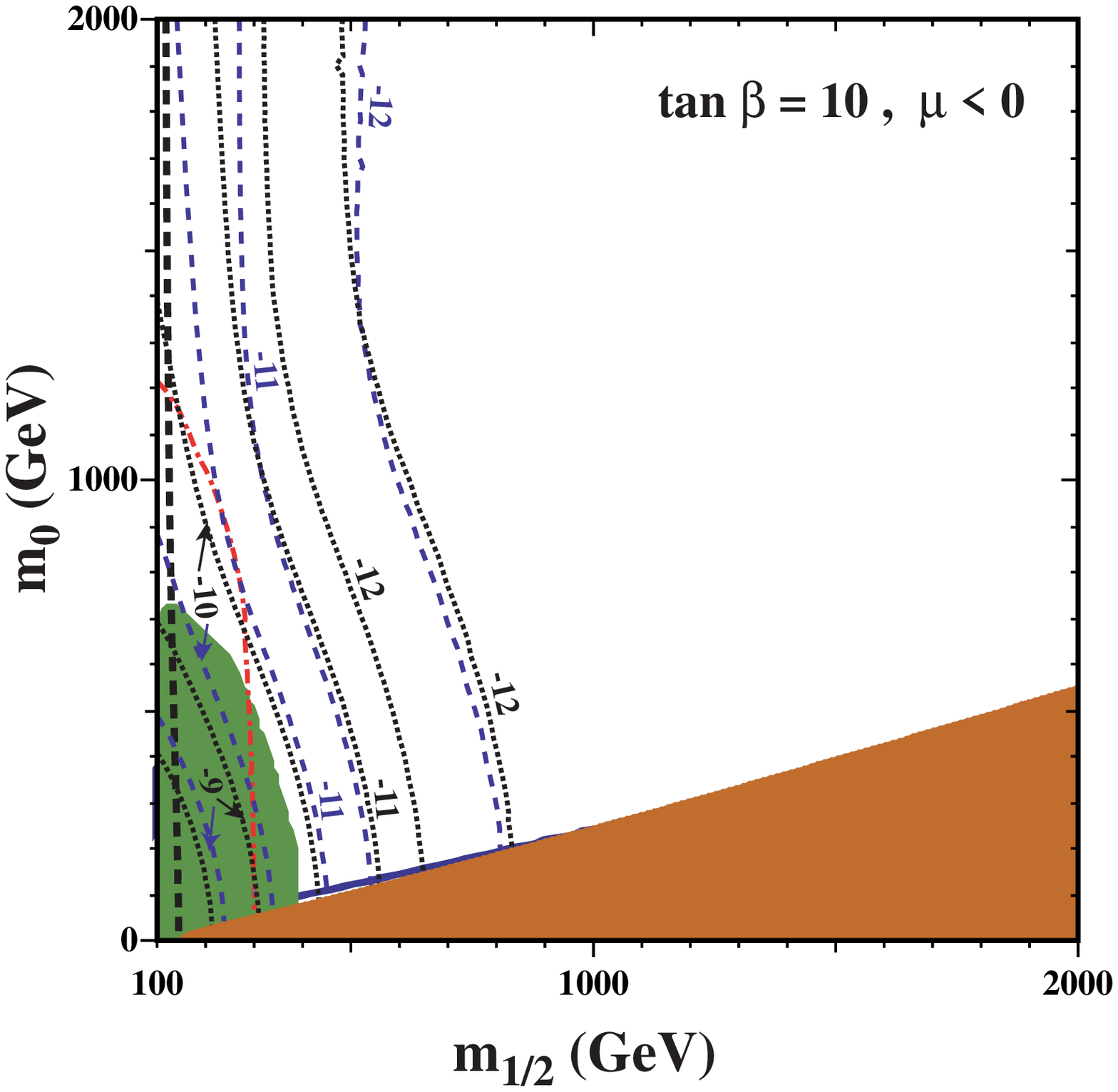,height=8cm}}
\mbox{\epsfig{file=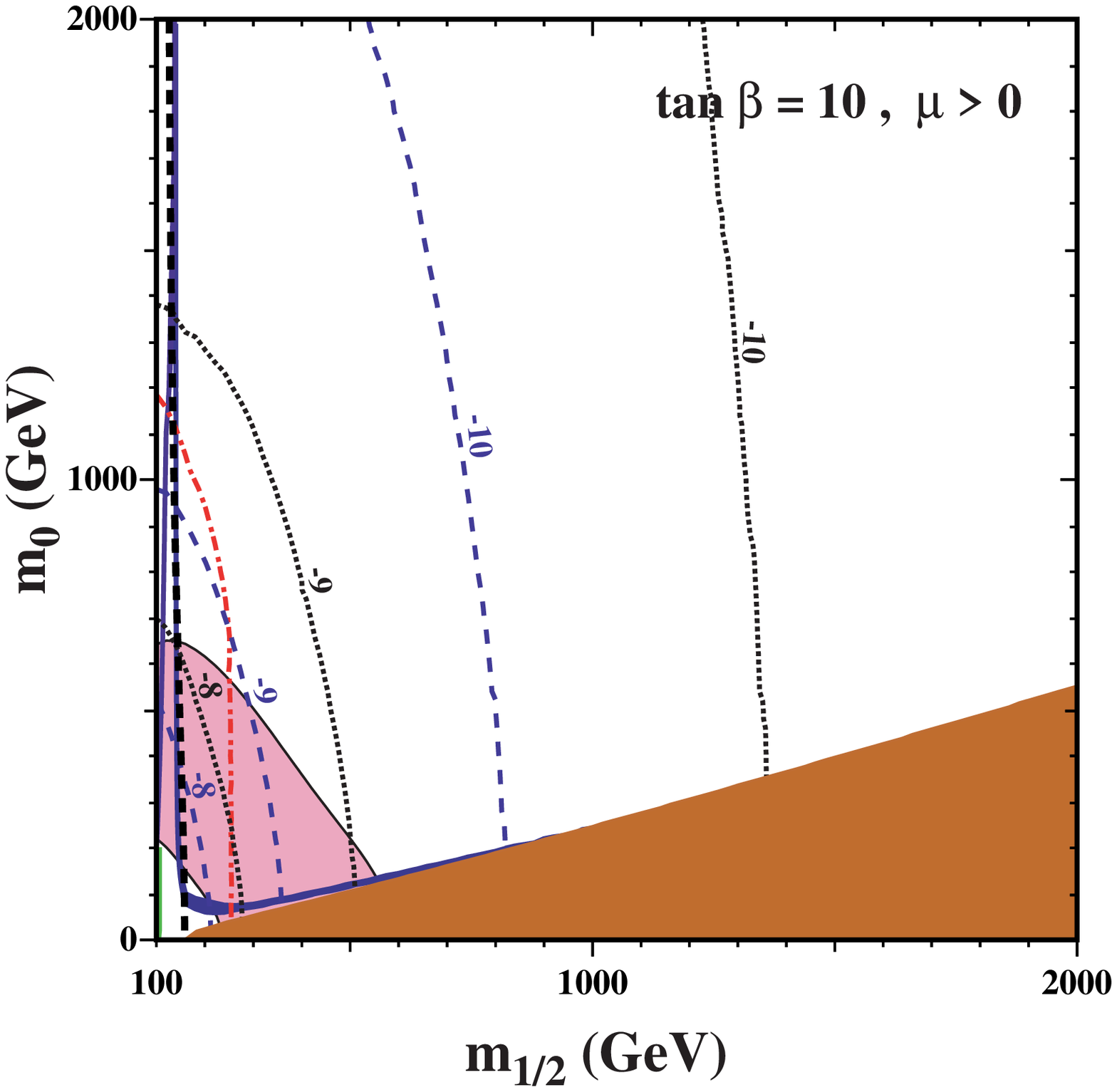,height=8cm}}
\end{center}
\begin{center}
\mbox{\epsfig{file=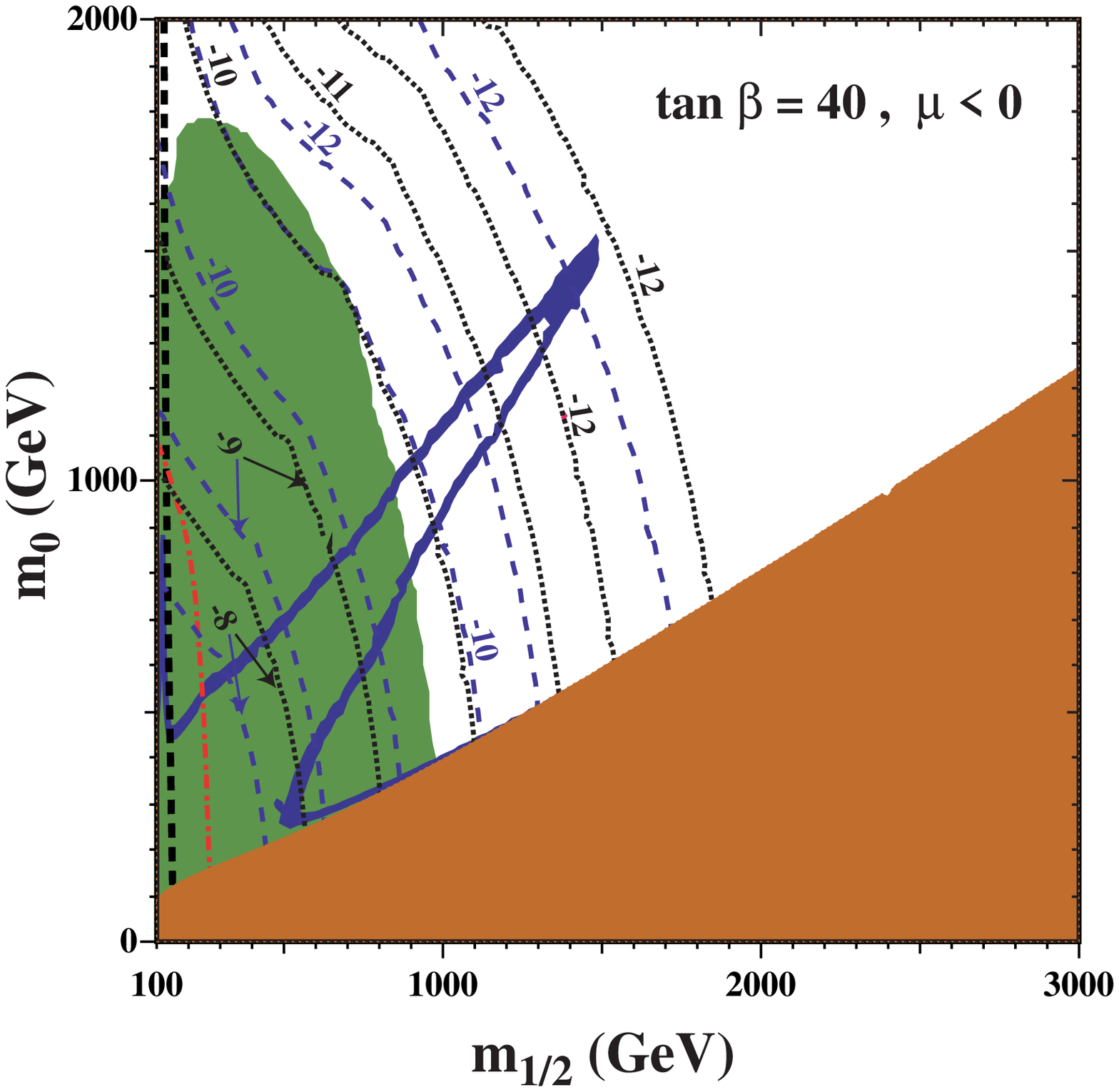,height=8cm}}
\mbox{\epsfig{file=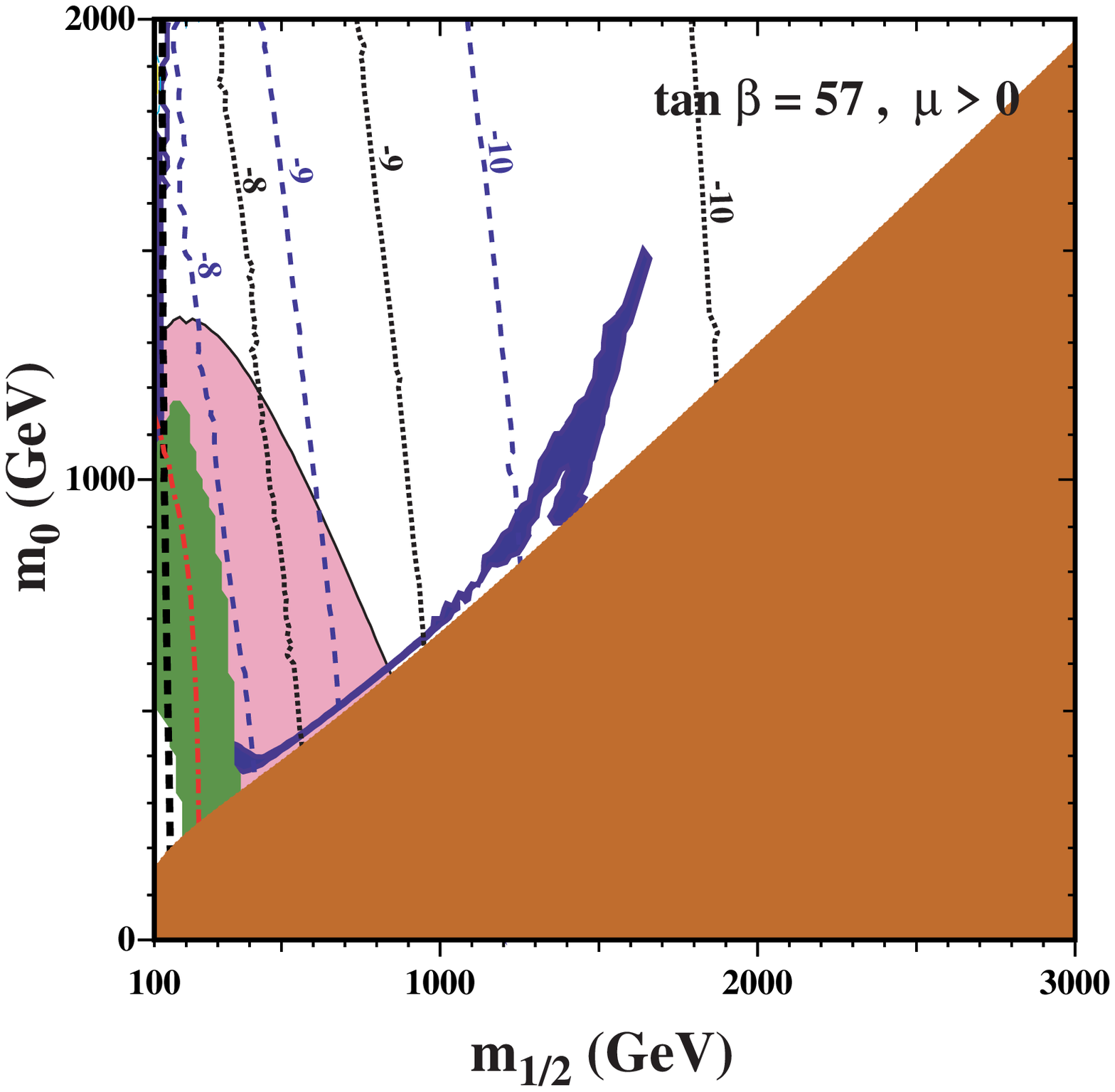,height=8cm}}
\end{center}
\caption{\label{fig:m0m12}
{\it 
The $(m_{1/2}, m_0)$ planes in the CMSSM for (a) $\tan \beta = 10, \ \mu <
0$, (b) $\tan \beta = 10, \ \mu > 0$, (c) $\tan \beta = 40, \ \mu < 0$ and (d)
$\tan \beta = 57, \ \mu > 0$, all assuming $A_0 = 0$. We display the WMAP 
relic-density constraint,
the experimental constraints due to $m_h$, $m_{\chi^\pm}$, $b \to s
\gamma$ and $g_\mu - 2$, and contours of the spin-independent 
elastic-scattering 
cross
section calculated for $\Sigma = 45$ and $64$~MeV (lighter, blue
and black dotted contours, respectively), labelled by 
their exponents in units of picobarns.
}} 
\end{figure}

The various experimental and cosmological constraints on the CMSSM are
displayed in various $(m_{1/2}, m_0)$ planes in Fig.~\ref{fig:m0m12}, but
we do not use them all as absolute limits. The dark, tan-shaded regions 
are, however, completely excluded
because there the LSP is charged, being the lighter ${\tilde \tau}$. The
thin blue strips are those favoured by the WMAP constraint on the relic
density of cold dark matter: $0.094 < \Omega_{CDM} h^2 < 0.125$ if
$\Omega_\chi \simeq \Omega_{CDM}$, and we also display the restrictions
that the accelerator constraints due to $m_h$ (red dash-dotted lines),
$m_{\chi^\pm}$ (black dashed lines) and $b \to s \gamma$ (medium, green 
shading)
impose on the ranges of $m_{1/2}$ and hence $m_0$ allowed along the WMAP
strips. The constraints that would be imposed by $g_\mu - 2$ at the
2-$\sigma$ level if the Standard Model contribution is evaluated using
$e^+ e^-$ annihilation data alone, neglecting $\tau$ decay data,
are shown by light, pink shading in panels (b, 
d)~\footnote{See~\cite{EHOW3} for a
discussion on the $g_\mu - 2$ deviation range used here. We recall that no 
models with $\mu < 0$ would be allowed at this significance level.}. 

Each of the panels also displays contours of the spin-independent
elastic-scattering cross section calculated for $\Sigma = 45$ (lighter,
blue dashed contours) and 64~MeV (black dotted contours), 
labelled by their exponents in units of picobarns. We see
that, for $\mu > 0$ in panels (b, d) of Fig.~\ref{fig:m0m12}, the
cross-section contours progress monotonically downward as $m_{1/2}$
increases, with the $\Sigma = 45$~MeV contours always at smaller $m_{1/2}$
than the corresponding contours for $\Sigma = 64$~MeV. However, the
progression is not monotonic for $\mu < 0$, as seen in panels (a, c). This
is because of the possibility of a cancellation between different
contributions to the scattering amplitude \cite{EFlO1}.

For the purpose of this paper, we choose to treat the WMAP constraint as
an upper limit on $\Omega_\chi h^2 \equiv f_\chi \Omega_{CDM} h^2: f_\chi
\le 1$, thus allowing for another component of cold dark matter with a
fractional density $1 - f_\chi \ge 0$. In this case, the small regions of
the $(m_{1/2}, m_0)$ planes between the WMAP strips and the charged LSP
corners are also allowed. We can see in Fig.~\ref{fig:m0m12} that the
spin-independent elastic scattering cross section is very similar in the
underdense regions with $f_\chi < 1$, which lie below the WMAP strips and
above the charged dark matter region, and also those inside the
rapid-annihilation funnels for large $\tan \beta$.

Implementing the accelerator constraints, using the relic density allowed
by WMAP as an upper limit: $\Omega_\chi h^2 = f_\chi \Omega_{CDM} h^2$,
and rescaling the cross section by a factor $f_\chi$ if $f_\chi < 1$, so
as to account for the fact that neutralinos could constitute only a
fraction $f_\chi$ of the galactic halo and that there would be another
important local component of cold dark matter, we find the ranges for the
effective spin-independent elastic-scattering cross section shown in
Fig.~\ref{fig:CMSSM}. These ranges were obtained by statistical sampling
of the allowed regions of the CMSSM parameter spaces for the indicated
parameter values. The sampling was performed over values of $m_{1/2} =
0.1$ to 2~TeV, $m_0 = 0$ to 2~TeV, $\tan \beta = 2$ to 43 (58) for
$\mu < (>) 0$, and $A_0 = -3 $ to $+3 \, m_{1/2}$. 
Because of the rescaling and the
fact that regions with $f_\chi < 1$ have similar intrinsic cross sections
to regions with $f_\chi = 1$, the points with $\Omega_\chi$ in the range
of $\Omega_{CDM}$ favoured by WMAP generally appear at the top of the
allowed ranges. In general, the calculated cross sections lie below the
present CDMS~II upper limits, except for certain models with the smallest
values of $m_{1/2}$ that are allowed when $\tan \beta \sim 10$ and $\mu >
0$, if one uses $\Sigma = 64$~MeV.

\begin{figure}
\begin{center}
\mbox{\epsfig{file=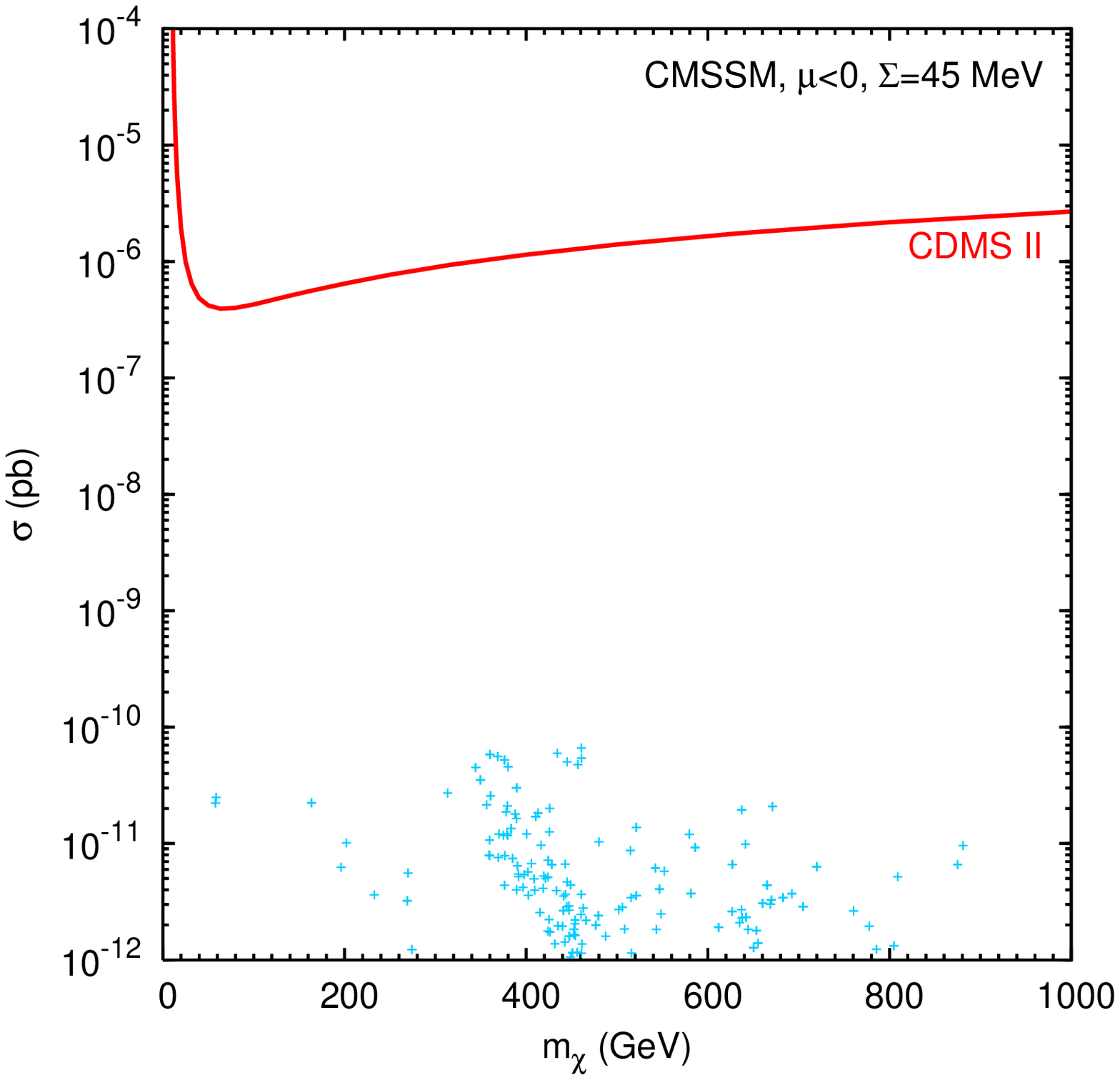,height=7cm}}
\mbox{\epsfig{file=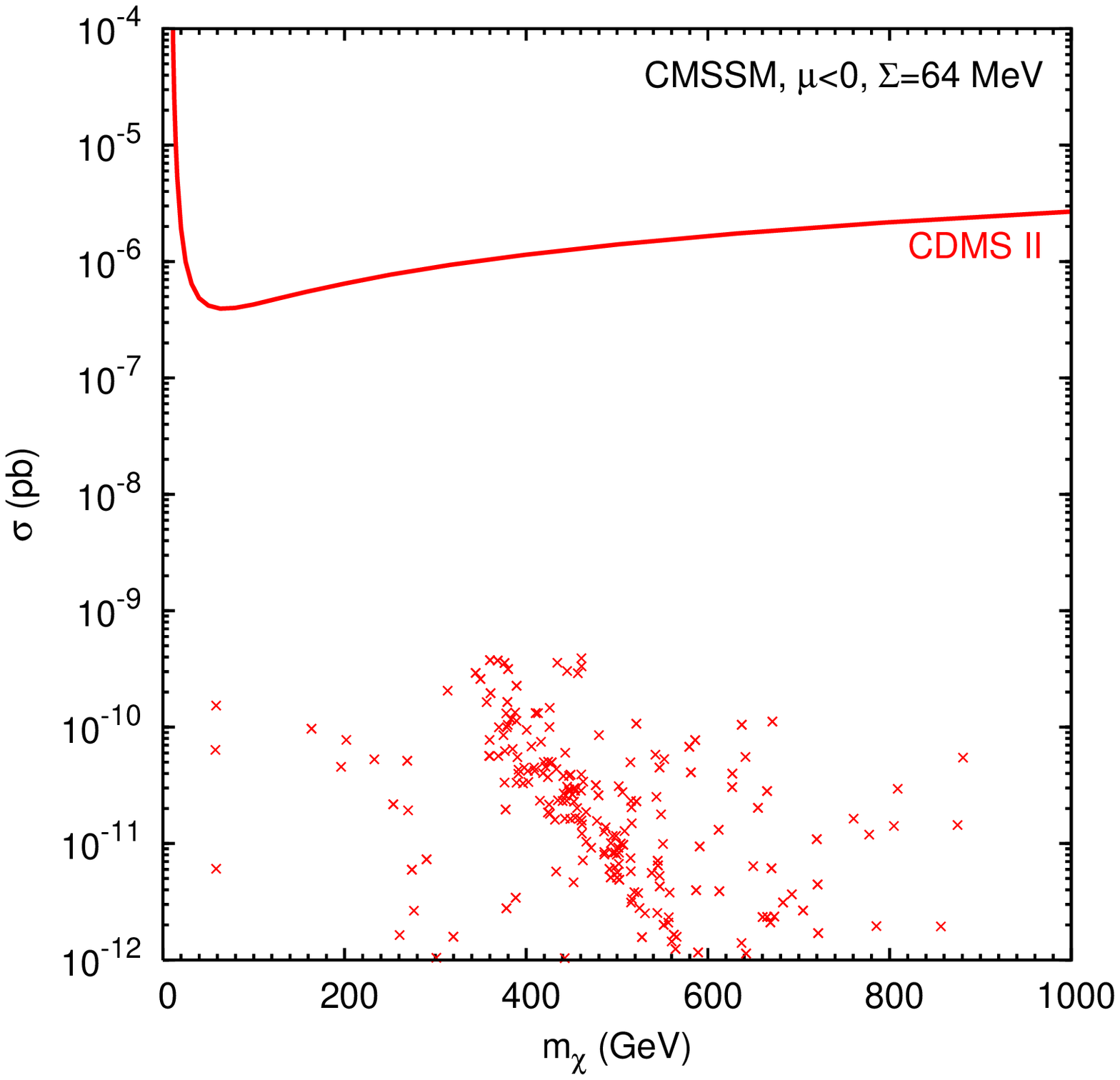,height=7cm}}
\end{center}
\begin{center}
\mbox{\epsfig{file=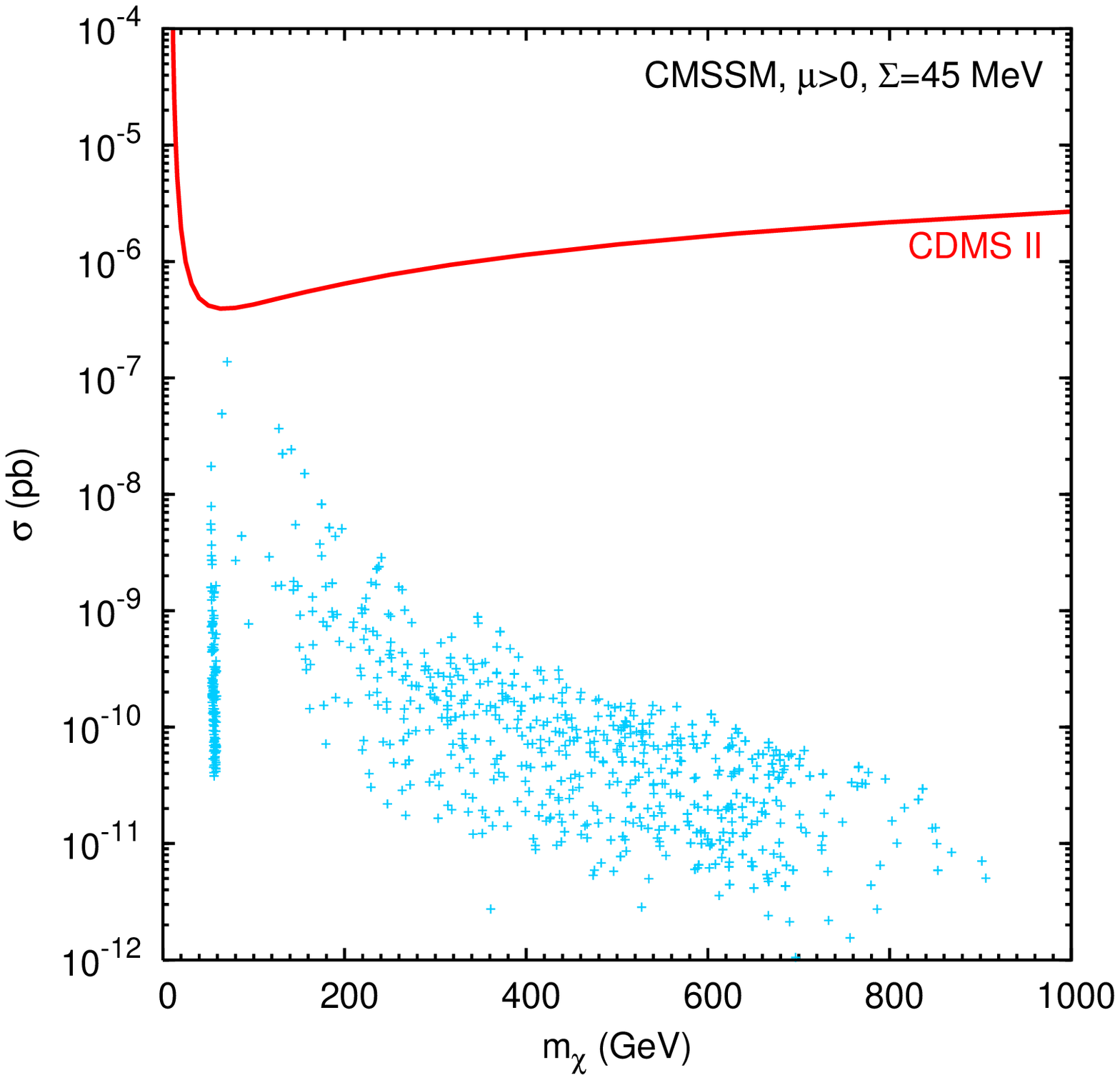,height=7cm}}
\mbox{\epsfig{file=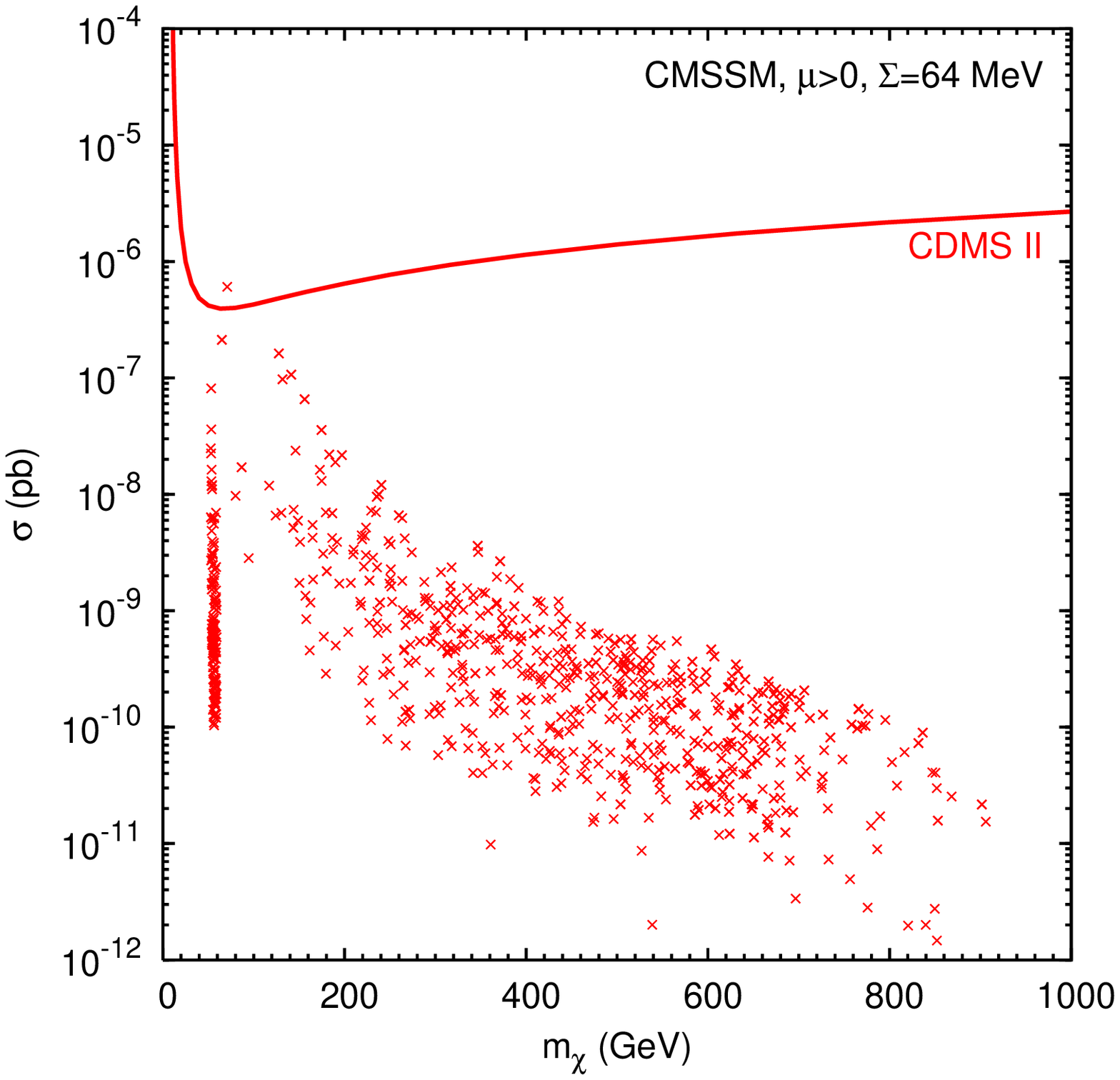,height=7cm}}
\end{center}
\caption{\label{fig:CMSSM}
{\it Scatter plots of the spin-independent elastic-scattering cross
section predicted in the CMSSM for (a, b) $\mu < 0$ and (c, d) $\mu > 
0$, with (a, c) $\Sigma = 45$~MeV and (b, d) $\Sigma = 64$~MeV.}}
\end{figure}

\subsection{Preferred Range of Sparticle Masses}

Progressing beyond the above implementation of laboratory experimental
constraints, the sparticle mass range preferred within the CMSSM has
recently been reassessed~\cite{EHOW3}, in light of recent precision
measurements and higher-order calculations in the Standard Model and the
MSSM. As has already been recalled, the anomalous magnetic moment of the
muon, $g_\mu - 2$, disagrees with the Standard Model by between 2.5 and 3
standard deviations~\cite{g-2}, if low-energy $e^+ e^-$ data are used to
estimate the strong-interaction contribution to $g_\mu - 2$: see the pink
shaded regions in Fig.~\ref{fig:m0m12}. The central experimental value
favours $\mu > 0$ and $m_{1/2} \sim 300$~GeV for $\tan \beta =
10$~\footnote{We note in passing that the minimum of the $\chi^2$ function
almost coincides with benchmark point B of~\cite{Bench2}, to which Point
1a of~\cite{sps} is also similar.}, and the preferred value of $m_{1/2}$
increases with $\tan \beta$. The present central values of $M_W$ and
$\sin^2 \theta_{eff}$ also disagree marginally with the latest theoretical
calculations within the Standard Model. Given the errors, these
discrepancies are not significant in themselves, but it so happens that
they are each, separately, best fit also by $m_{1/2} \sim 300$~GeV for
$\tan \beta = 10$. The quality of fit in the $(m_{1/2}, A)$ planes for
$\tan \beta = 10, 50$ has been explored, and the 68\% and 90\% confidence
level regions have been delineated: they stretch up to $m_{1/2} \lappeq
1000$~GeV~\cite{EHOW3}.

In Fig.~\ref{fig:EHOW3} we display scatter plots of the spin-independent
elastic-scattering cross section calculated for $\Sigma = 45$ and 64~MeV,
as usual, for the portions of the WMAP strips allowed for (a, b) $\tan
\beta = 10, \mu > 0$ and (c, d) $\tan \beta = 50, \mu > 0$ at both the
68\% and 90\% confidence levels. The two choices CL = 68 \% and 90 \% have
different colours (darker, blue $\times$ and lighter, green $+$ signs,
respectively).  We do not see large qualitative differences between the
cross-section predictions in the 68\% and 90\% confidence-level cases.
Also, comparing the top and bottom panels, we do not see large qualitative
differences between the cross-section predictions in the cases $\tan \beta
= 10,\ 50$, though the latter are slightly larger.  However, comparing the
left and right panels, we once again see the direct effect on the cross
section due to our choice of $\Sigma$.  Since $\mu > 0$ in this analysis,
there is no possibility of a cancellation in the cross section. Moreover,
comparing with the corresponding panels of Fig.~\ref{fig:CMSSM}, we note
that the preferred range of $m_{1/2}$ and hence $m_\chi$ happens to be
that where the spin-independent elastic scattering cross section is relatively
large.

\begin{figure}
\begin{center}
\mbox{\epsfig{file=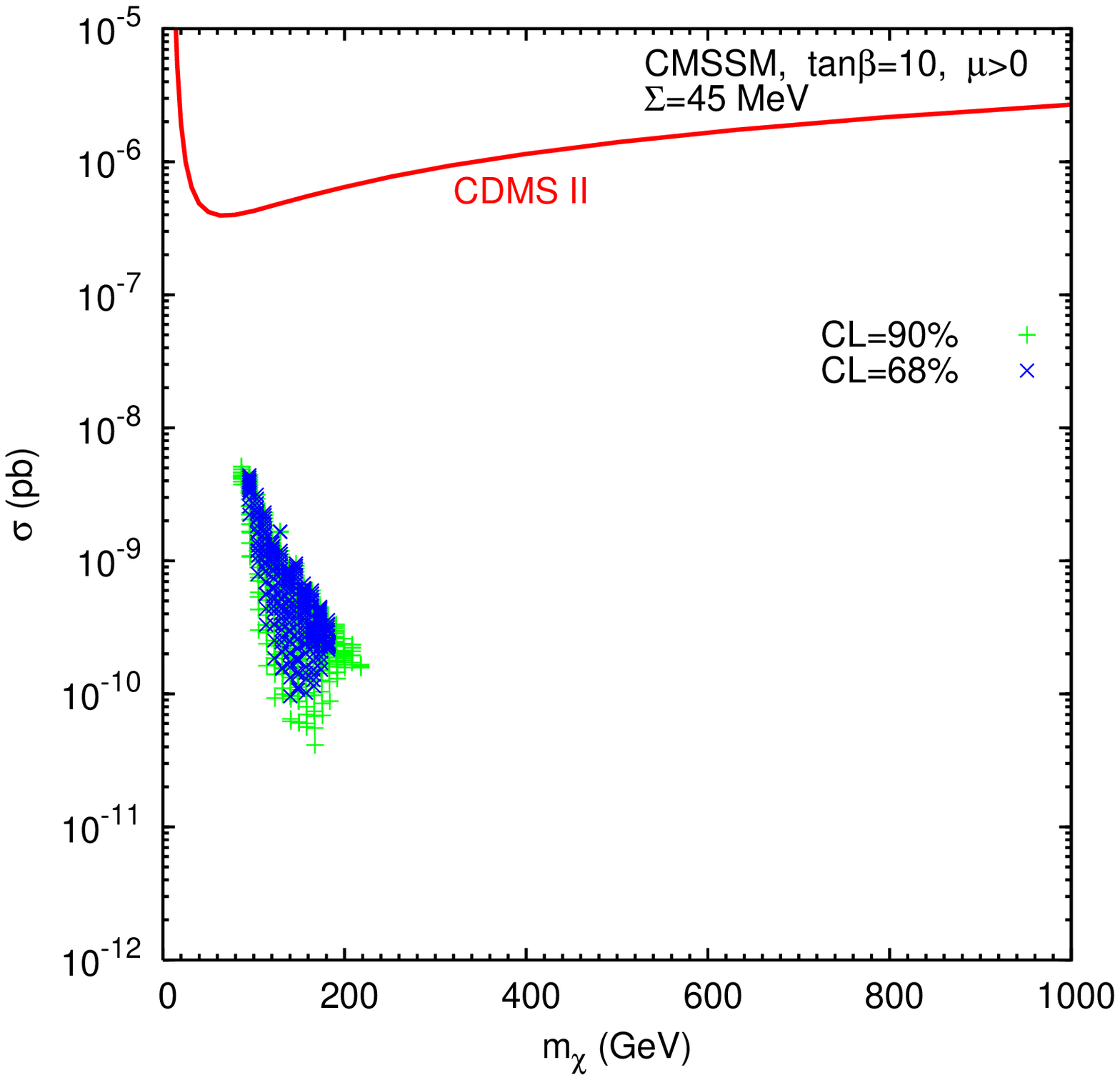,height=7cm}}
\mbox{\epsfig{file=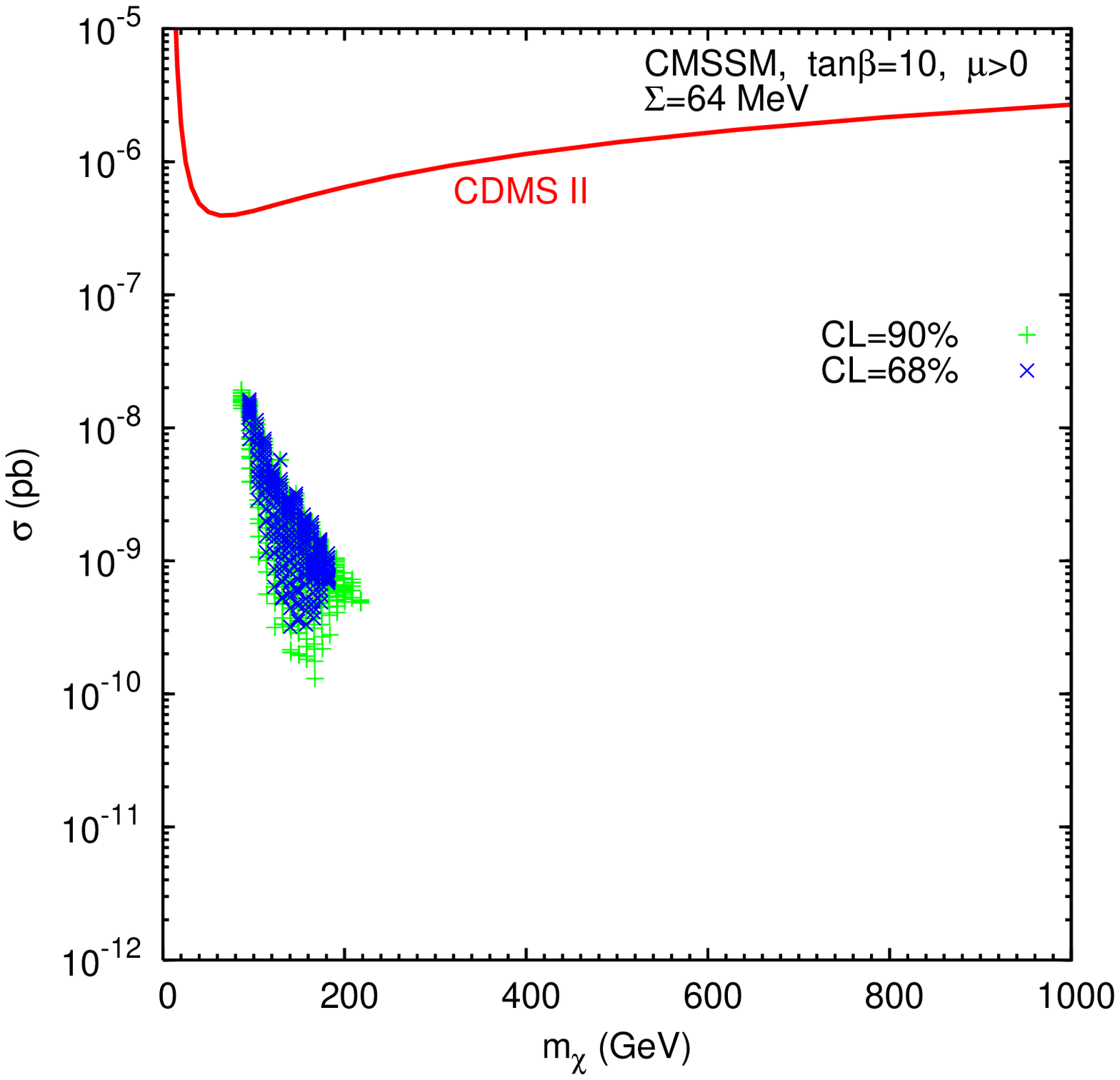,height=7cm}}
\end{center}
\begin{center}
\mbox{\epsfig{file=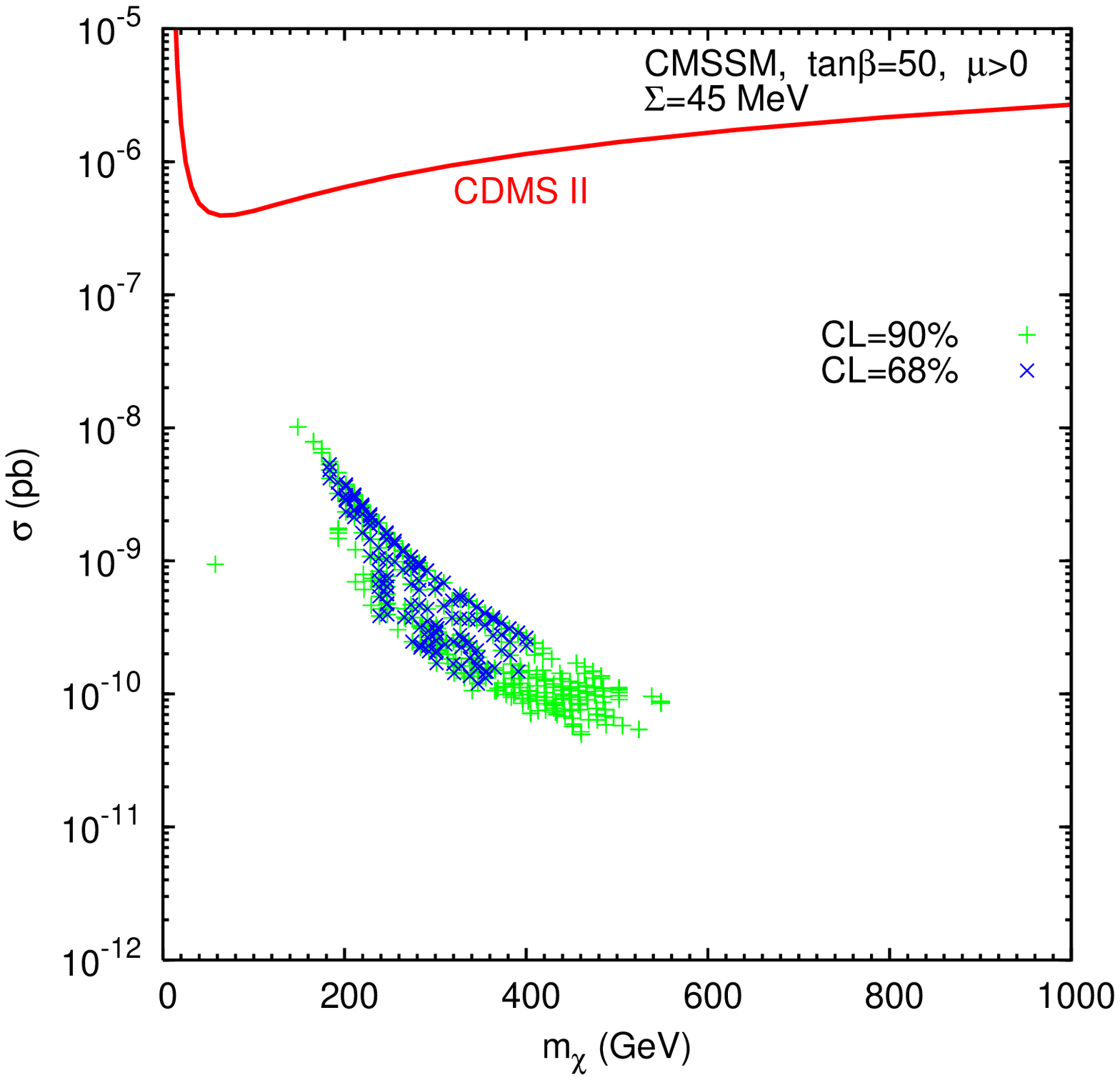,height=7cm}}
\mbox{\epsfig{file=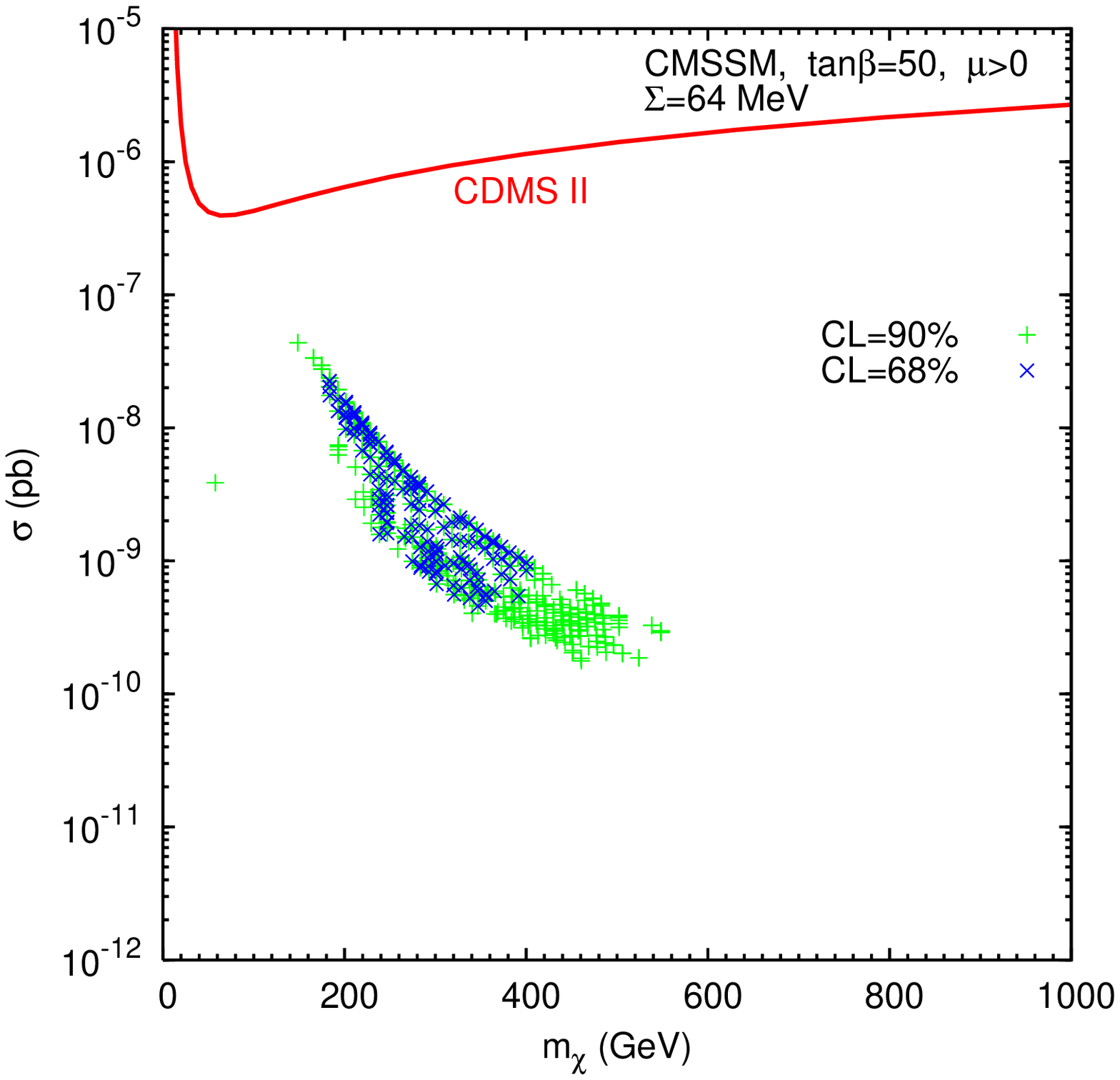,height=7cm}}
\end{center}
\caption{\label{fig:EHOW3}
{\it Scatter plots of the spin-independent elastic-scattering cross
section predicted in the CMSSM for (a, b) $\tan \beta = 10, \mu > 0$ and 
(c, d) $\tan \beta = 50, \mu > 0$, with (a, c) $\Sigma = 45$~MeV and (b, d) $\Sigma = 
64$~MeV. The predictions for models allowed at the 68\% (90\%) confidence 
levels are shown by blue $\times$ signs (green $+$ signs).}} 
\end{figure}

We see that an improvement in the present CDMS~II limit by an order of
magnitude would just begin to touch the estimated cross-section range, for
low $m_\chi$ and large $\Sigma$. On the other hand, an improvement by 
around 4 orders of magnitude would be required to cover completely all the 
regions allowed at the 90\% confidence level for all the considered range 
45~MeV$ < \Sigma <$64~MeV.

\section{Detection in Models with Non-Universal Scalar Masses}

Larger cross sections may be found in models in which the CMSSM
assumptions of universal soft supersymmetry-breaking scalar masses $m_0$
are relaxed, as we now discuss.

\subsection{General Low-Energy Effective Supersymmetric Theory}

We first consider relaxing the universality assumption for the Higgs
bosons and for the soft supersymmetry-breaking squark masses relative to
those of the sleptons, requiring only that all the squark and slepton
squared masses remain positive under renormalization up to the GUT scale.
This we term the most general low-energy effective supersymmetric theory
(LEEST)~\cite{LEEST}.  It is clear that relaxing the CMSSM relationship
between the squark and slepton masses $m_{\tilde q}, m_{\tilde \ell}$
might have a direct impact on the elastic-scattering cross section,
although the freedom to adjust $m_{\tilde q} / m_{\tilde \ell}$ is
severely restricted by the LEEST requirement that the squared masses
remain positive up to the GUT scale. The primary impact of relaxing
universality for the Higgs boson masses is to permit variations from the
CMSSM values of the pseudoscalar Higgs mass $m_A$ and the magnitude of
Higgs mixing $|\mu|$, which are fixed by the electroweak vacuum
conditions. We discuss below the extent to which these different effects
can be disentangled.
 
We display in Fig.~\ref{fig:LEEST} scatter plots of the spin-independent
elastic-scattering cross section for both signs of the Higgs-mixing
parameter $\mu$: negative in panels (a, b) and positive in panels (c, d).  
Predictions for two values of $\Sigma$, the conservative value of 45~MeV
and the more modern value of 64~MeV, are shown in panels (a, c) and (b,
d), respectively. We see that predictions of $\Sigma$ for $\mu < 0$ never
rise to the sensitivity of the CDMS~II experiment~\cite{CDMS2}, whichever
value of $\Sigma$ is used.  However, a few points with $m_\chi \lappeq
700$~GeV do exceed the current CDMS~II limit for $\mu > 0$, as seen in
panels (c, d), particularly when the larger value of $\Sigma$ is used. We discuss
the nature of these excluded points further below.

\begin{figure}
\begin{center}
\mbox{\epsfig{file=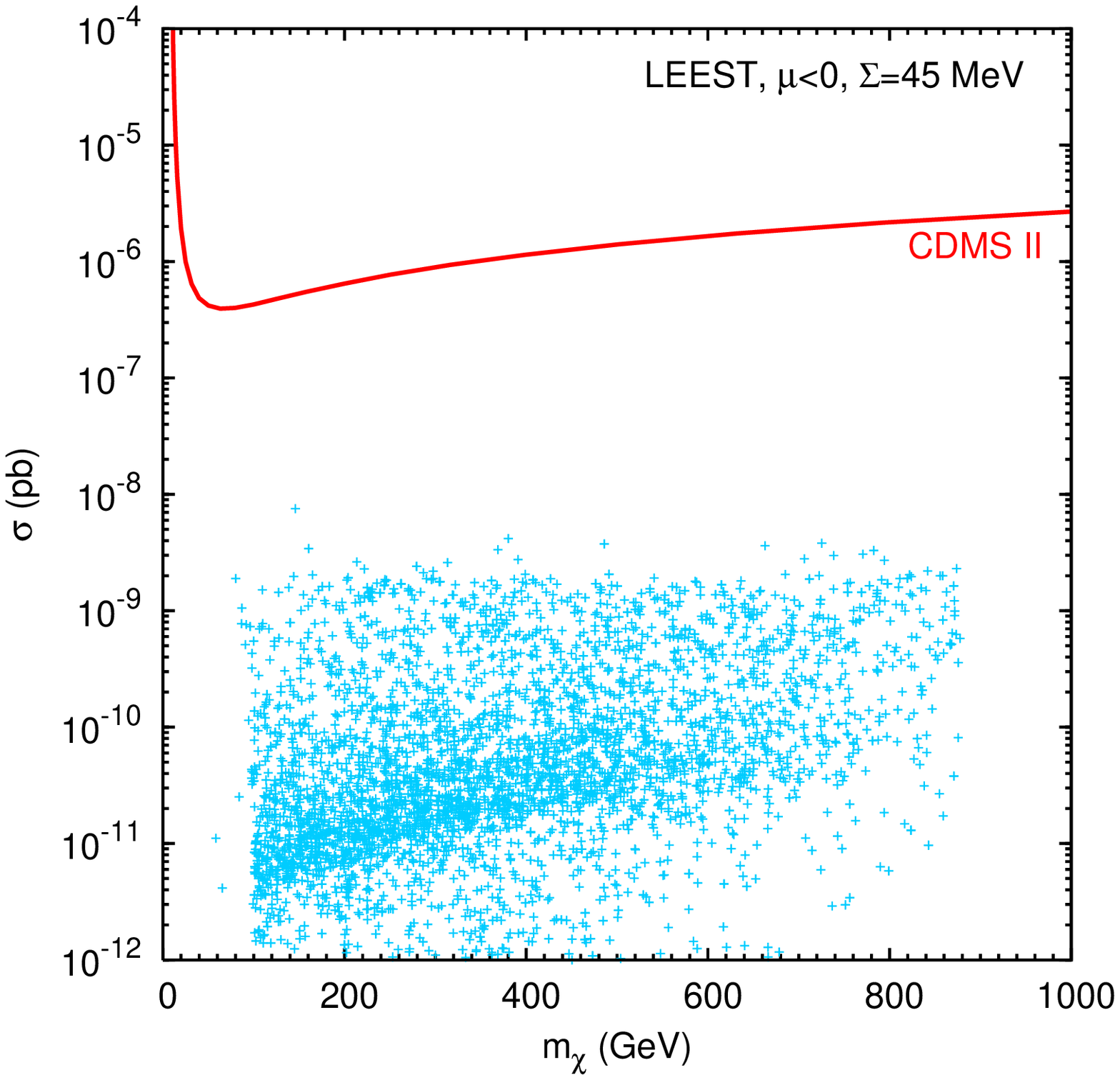,height=7cm}}
\mbox{\epsfig{file=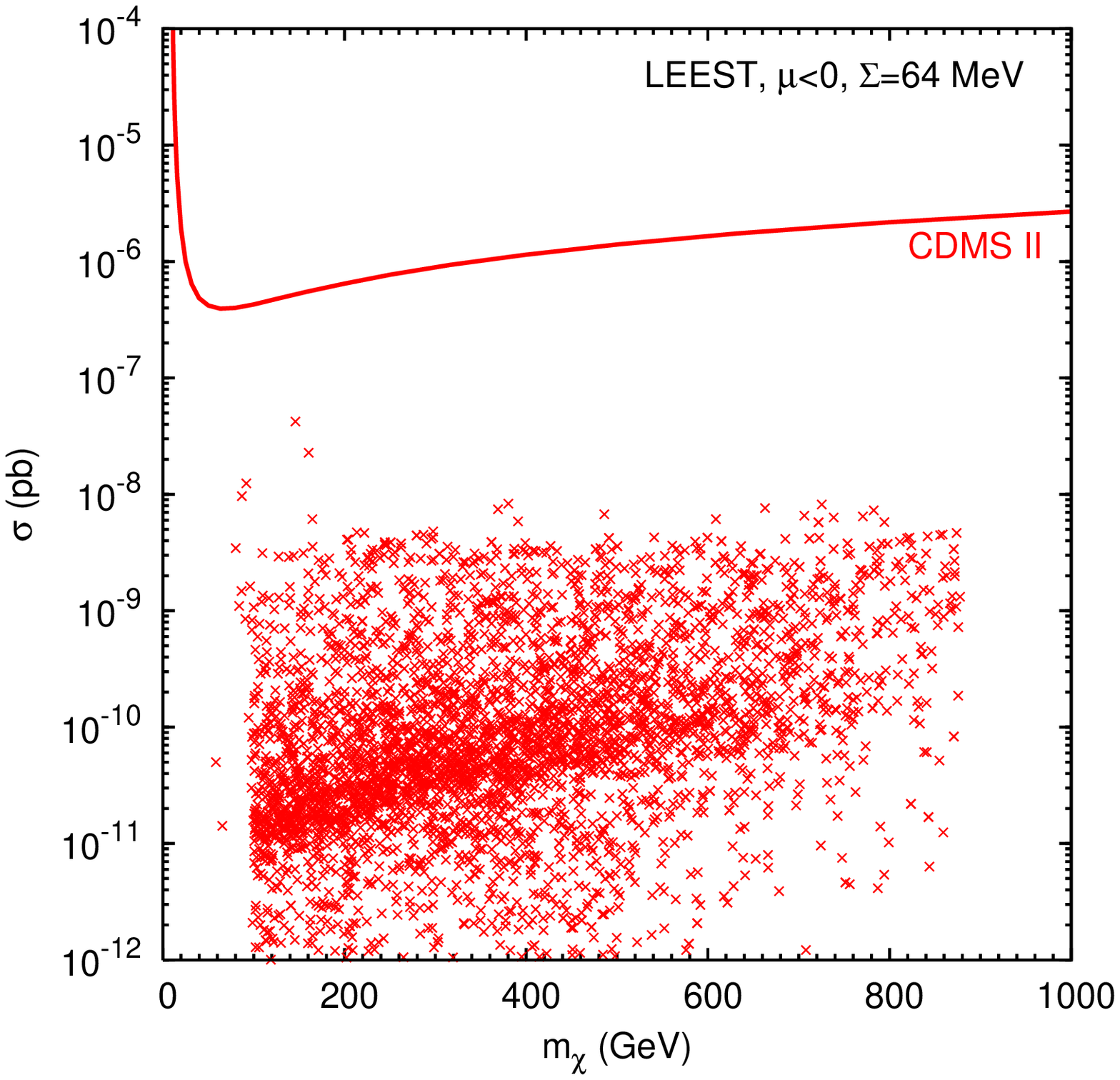,height=7cm}}
\end{center}
\begin{center}
\mbox{\epsfig{file=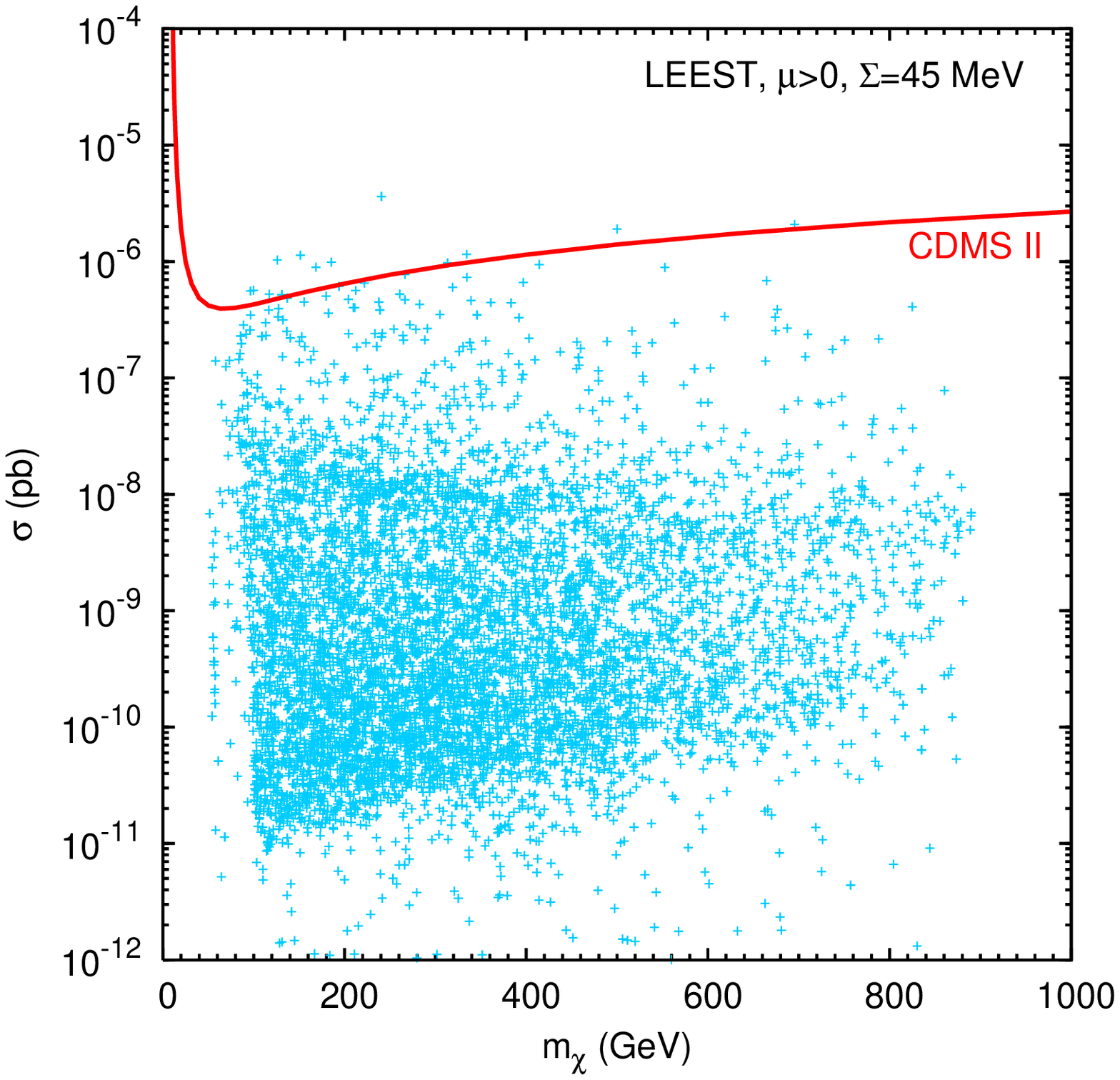,height=7cm}}
\mbox{\epsfig{file=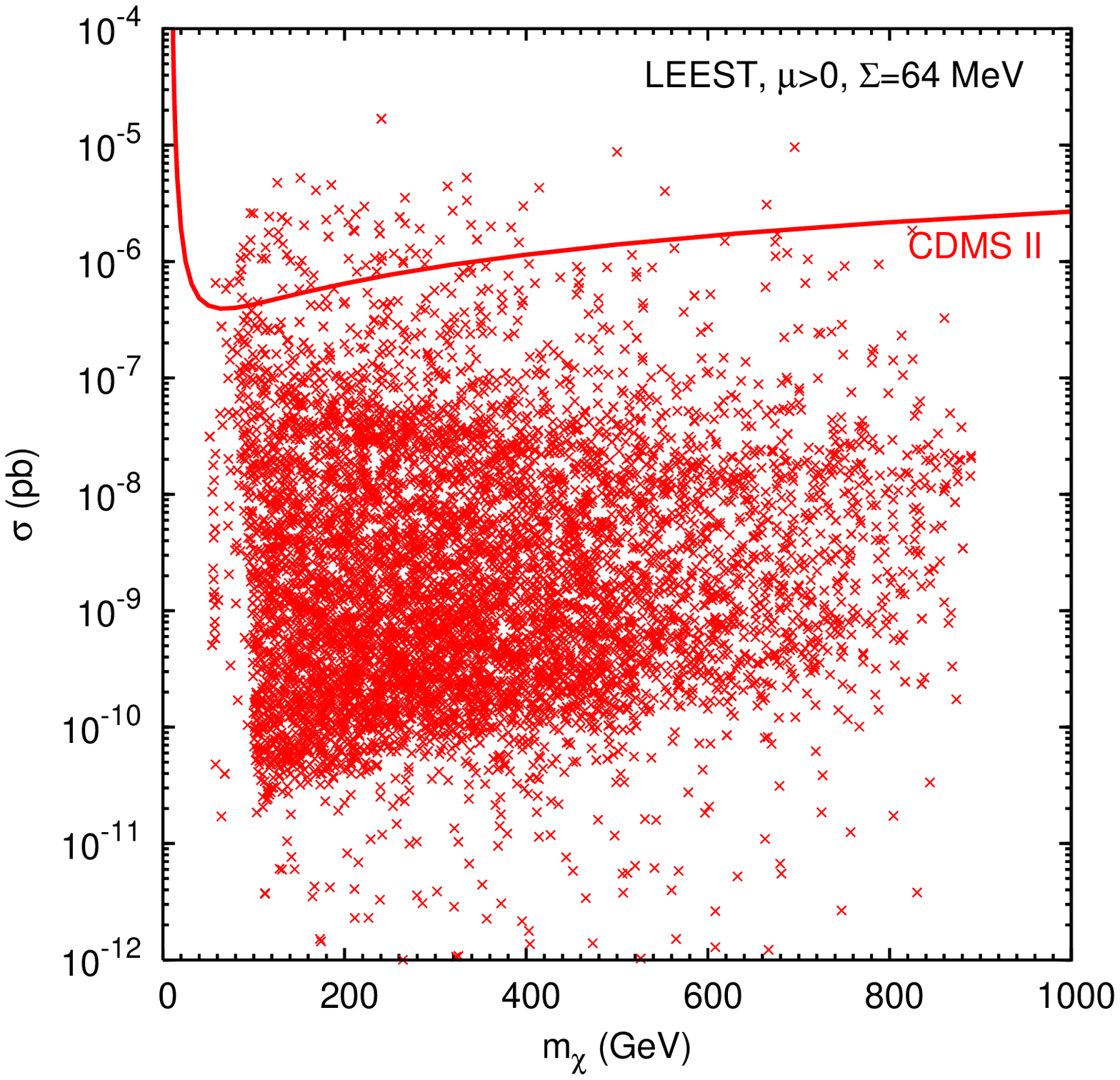,height=7cm}}
\end{center}
\caption{\label{fig:LEEST}
{\it As in Fig.~\protect\ref{fig:CMSSM}, but now for the LEEST.}}
\end{figure}

\subsection{Models with Non-Universal Higgs Masses}

Since the parameter space of the LEEST has quite a large dimensionality,
it is difficult to visualize clearly what classes of models might be
excluded by CDMS~II. This becomes clearer if one considers a class of
models with a lower-dimensional parameter space, namely those with
universal soft supersymmetry-breaking masses for squarks and sleptons but
non-universal Higgs masses (NUHM)~\cite{nonu,nuhm,EFlOSo}, which allow 
values of $|\mu|$ and $m_A$ differing from those in the CMSSM. We display in
Fig.~\ref{fig:NUHM} scatter plots of the spin-independent
elastic-scattering cross section for (a, b) $\mu < 0$ and (c, d) $\mu >
0$. The same two choices of $\Sigma$, namely 45~MeV and 64~MeV are made in
panels (a, c) and (b, d), respectively.

\begin{figure}
\begin{center}
\mbox{\epsfig{file=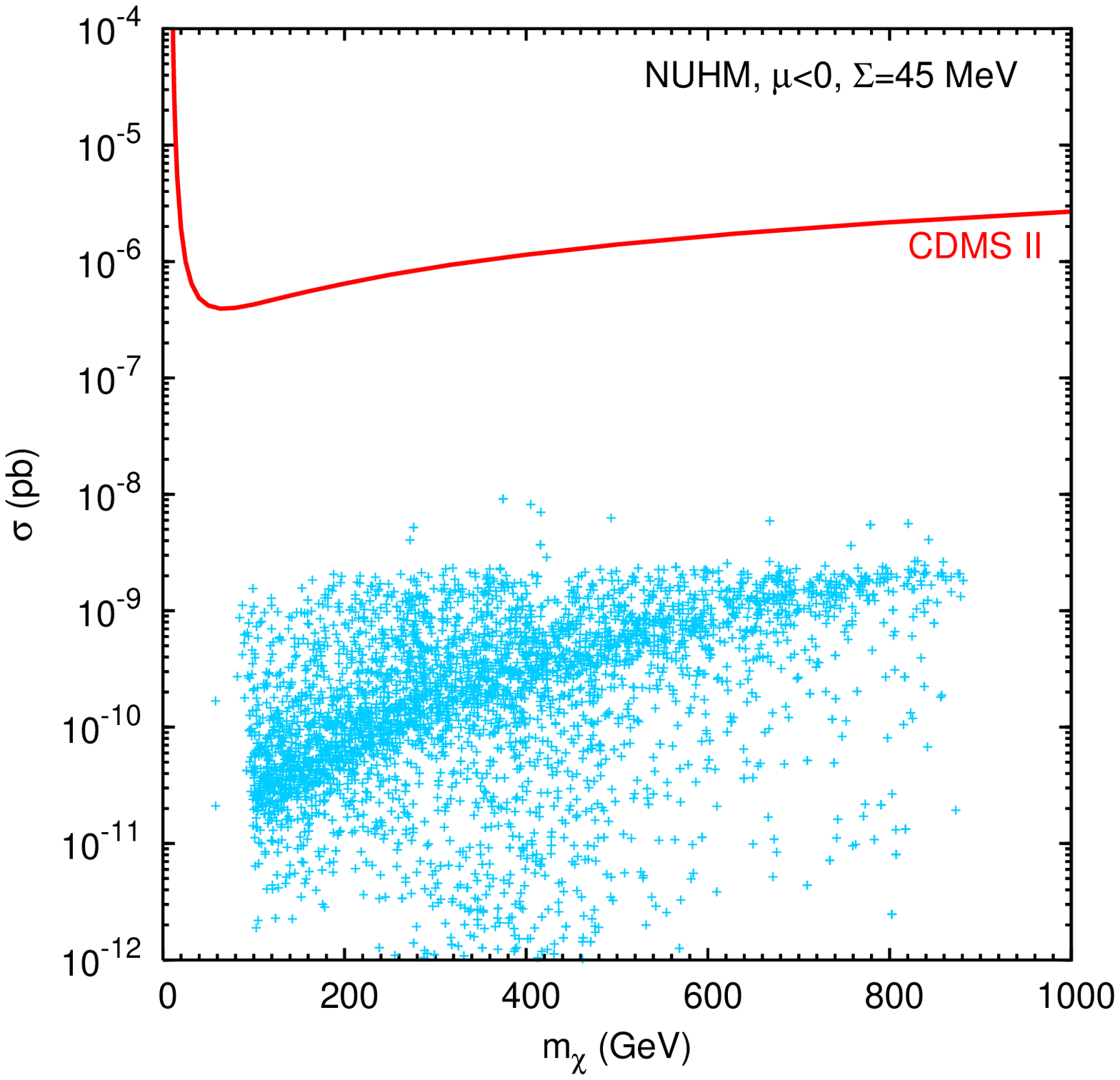,height=7cm}}
\mbox{\epsfig{file=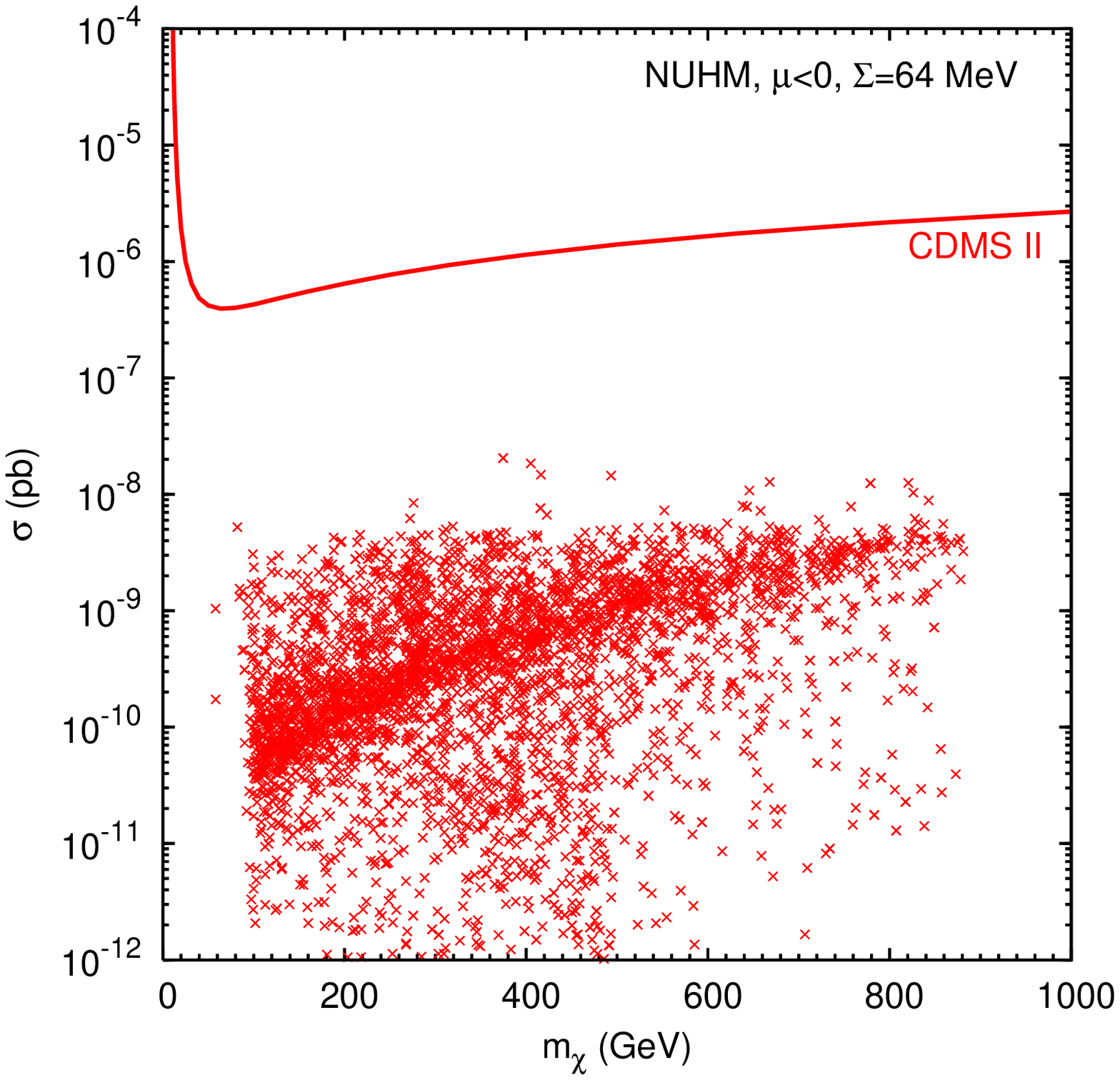,height=7cm}}
\end{center}
\begin{center}
\mbox{\epsfig{file=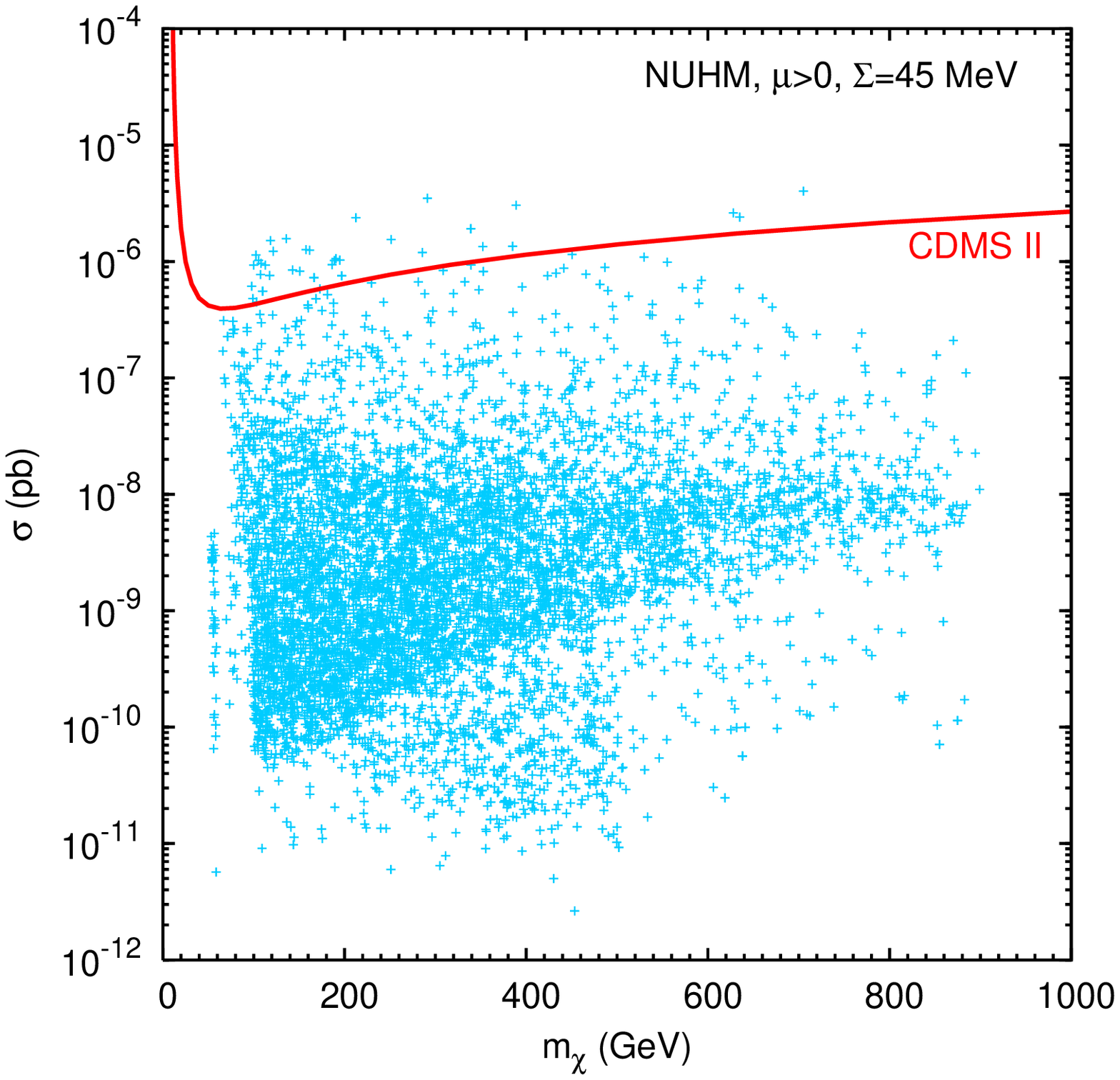,height=7cm}}
\mbox{\epsfig{file=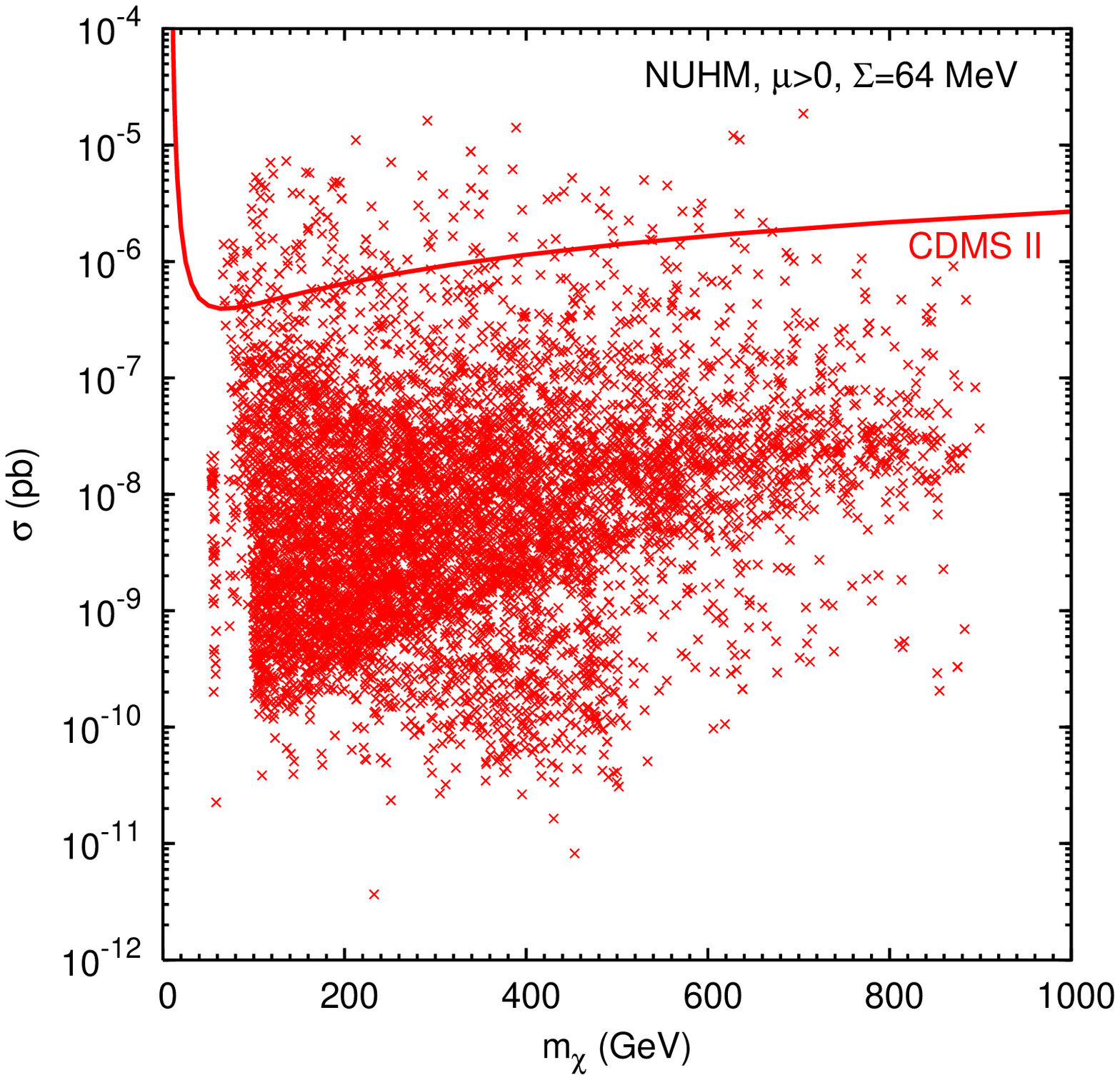,height=7cm}}
\end{center}
\caption{\label{fig:NUHM}
{\it As in Fig.~\protect\ref{fig:LEEST}, but now for the NUHM. }}
\end{figure}

We see again that no NUHM points can be excluded for $\mu <0$ but that, as
in the LEEST case, some $\mu > 0$ NUHM points may be excluded by CDMS~II.
This is true, in particular, for the larger choice of $\Sigma$. The
similarities between the general trends in the corresponding panels of
Figs.~\ref{fig:LEEST} and \ref{fig:NUHM} suggest that the dominant
effects may be due to relaxing the universality assumption for the Higgs
masses which, we recall, allows the values of $|\mu|$ and $m_A$ to differ
from those in the CMSSM. In fact, the LEEST does not have much leeway for
varying the ratio $m_{\tilde q}/m_{\tilde \ell}$ at low energies, since we
restrict the soft supersymmetry-breaking scalar masses so that the
effective scalar squared masses remain non-tachyonic all the way up the
GUT scale.

The fact that the most significant variations from the CMSSM are likely to
be those in $|\mu|$ and $m_A$ is supported by a previous general study of
the NUHM~\cite{nuhm}, in which various $(m_{1/2}, m_0)$, $(m_A, \mu)$ and
$(m_A, M_2)$ planes were exhibited. The behaviours of the cross section in
the NUHM $(m_{1/2}, m_0)$ planes were similar to those found in the CMSSM,
varying mainly with $m_{1/2}$ and less with $m_0$ \cite{EFlOSo}. The dependence on $M_2$
in the $(m_A, M_2)$ planes basically reflected the same $m_{1/2}$
dependence. The most striking dependence of the cross section was on
$|\mu|$, so we focus here on the $(m_A, \mu)$ planes for $m_{1/2} =
500$~GeV, $m_0 = 1000$~GeV and different choices of $\tan \beta $, which
are displayed in Fig.~\ref{fig:CDMS2}. Regions outside and below the black
double-dash-dotted lines have negative Higgs masses-squared below the GUT 
scale,
and are hence unstable, so only the regions between and above these lines
are allowed. This constraint becomes less important as $\tan \beta$ is 
increased.

\begin{figure}
\begin{center}
\mbox{\epsfig{file=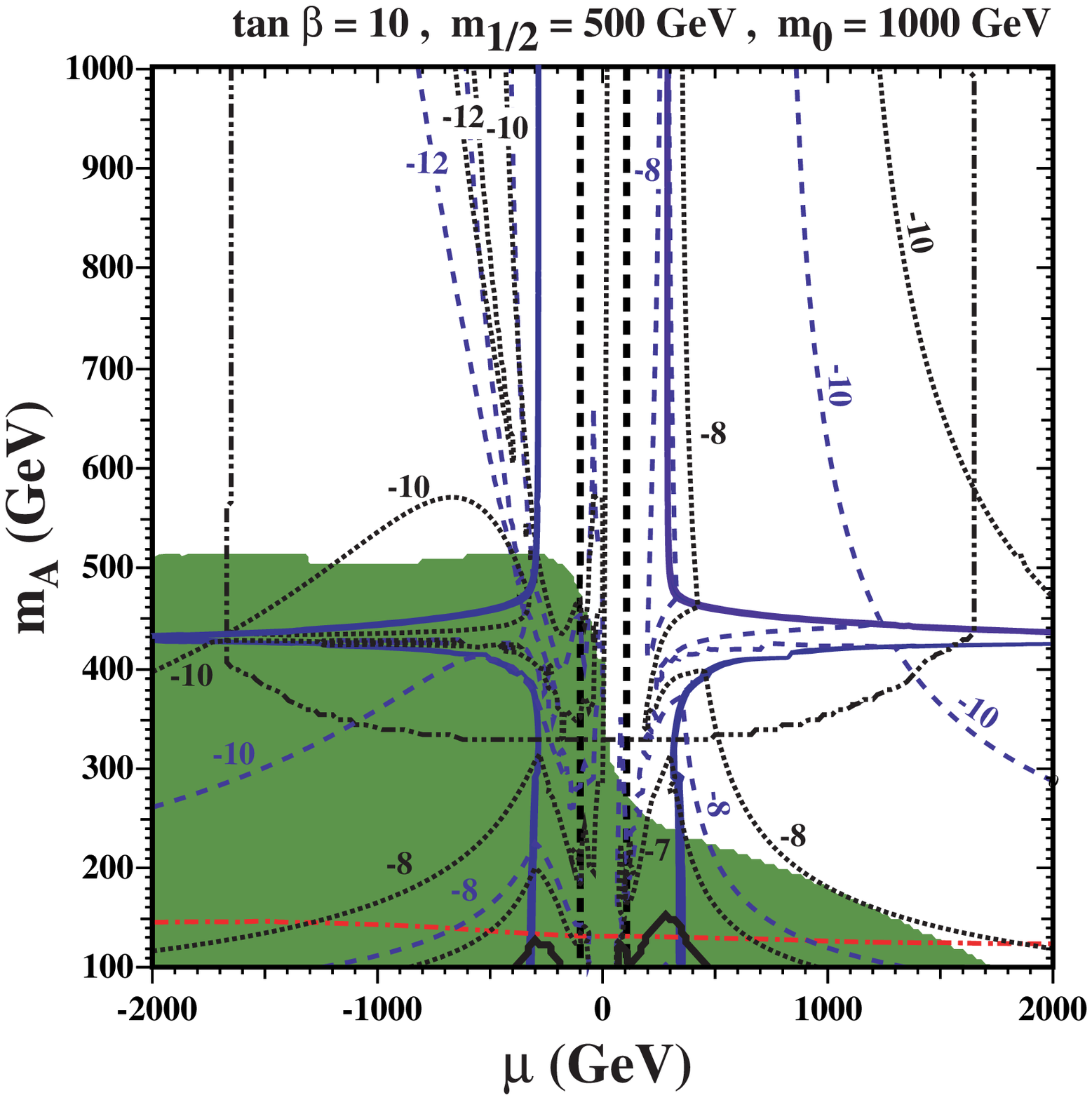,height=8cm}}
\mbox{\epsfig{file=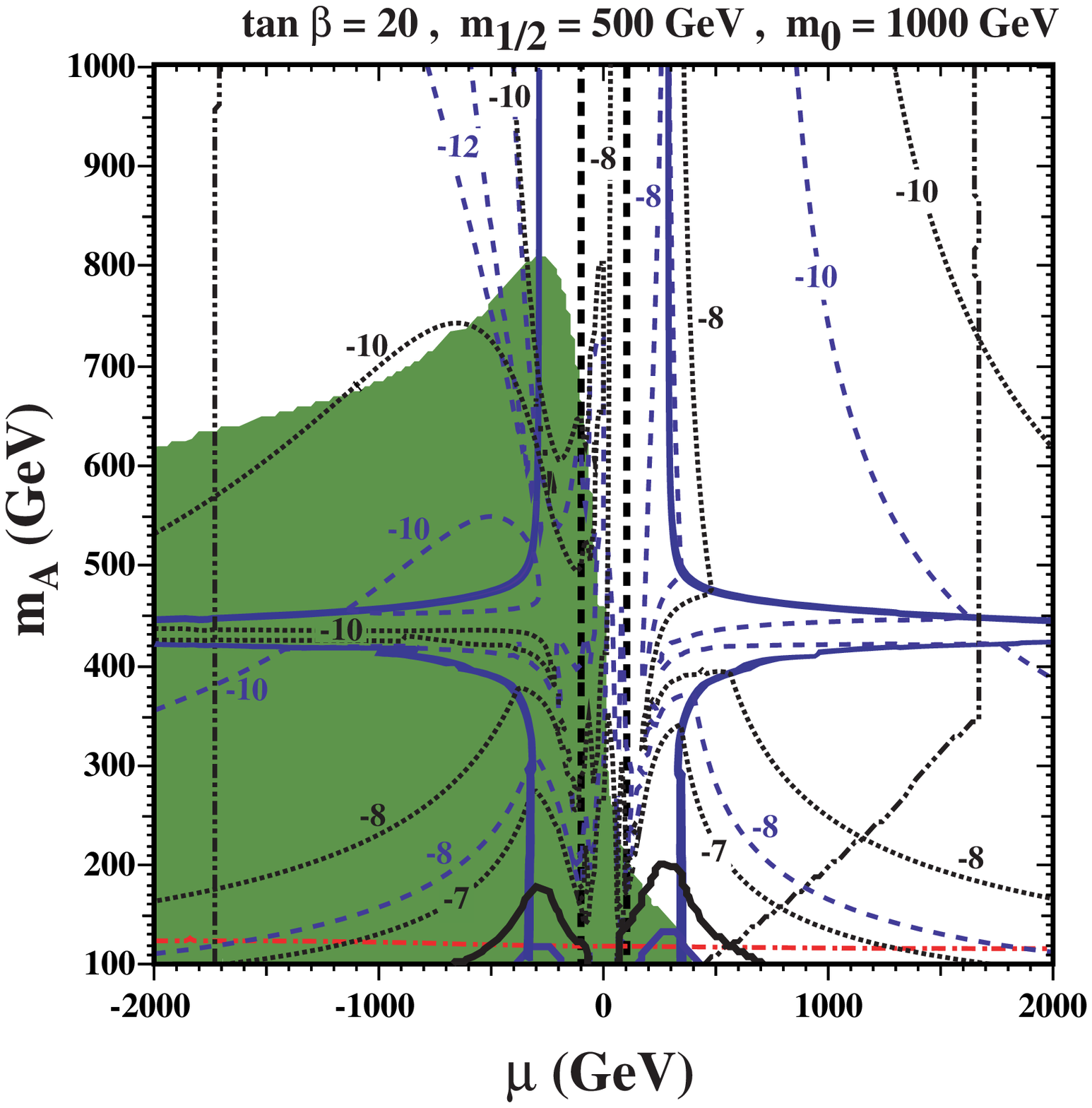,height=8cm}}
\end{center}
\begin{center}
\mbox{\epsfig{file=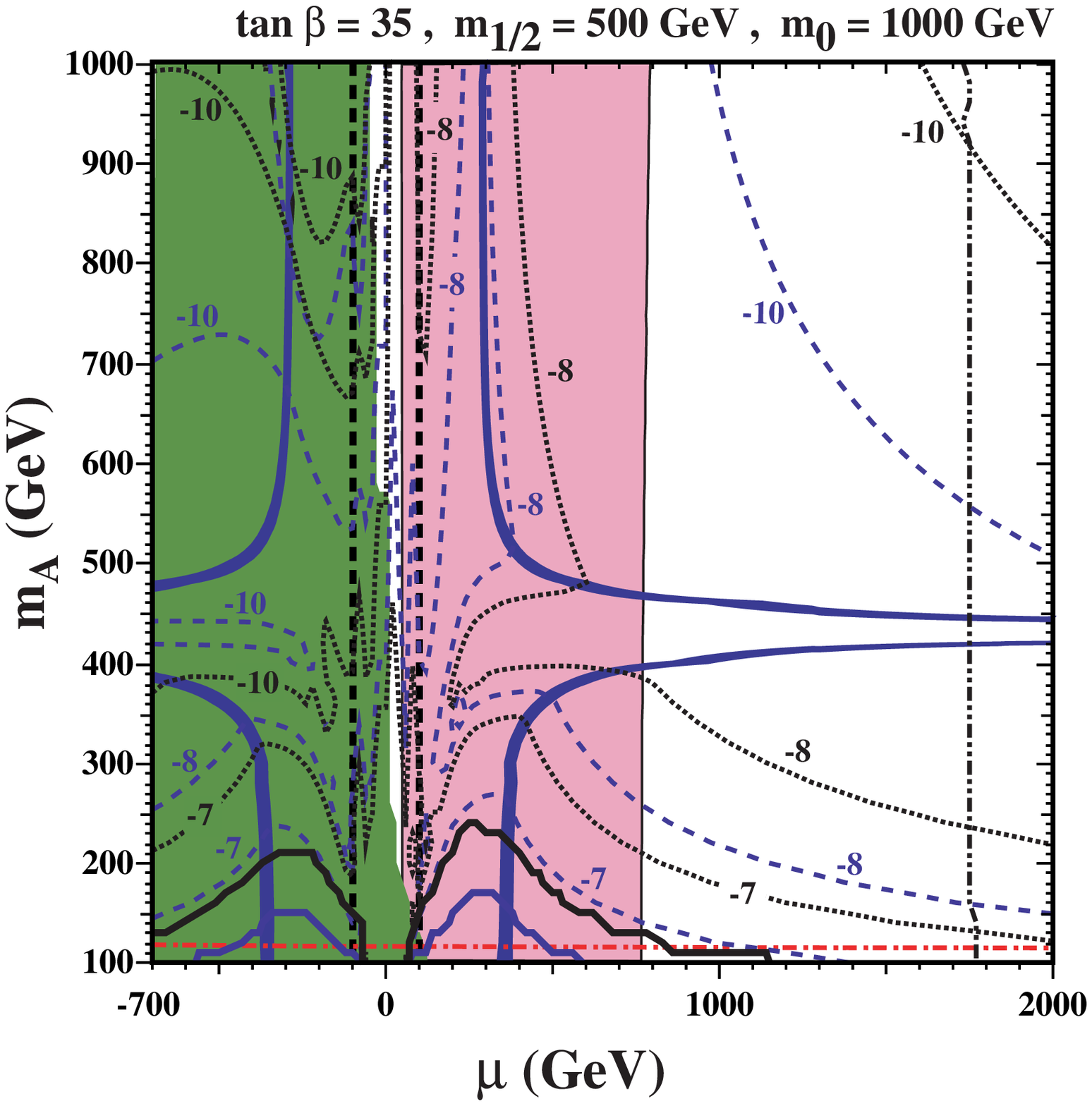,height=8cm}}
\mbox{\epsfig{file=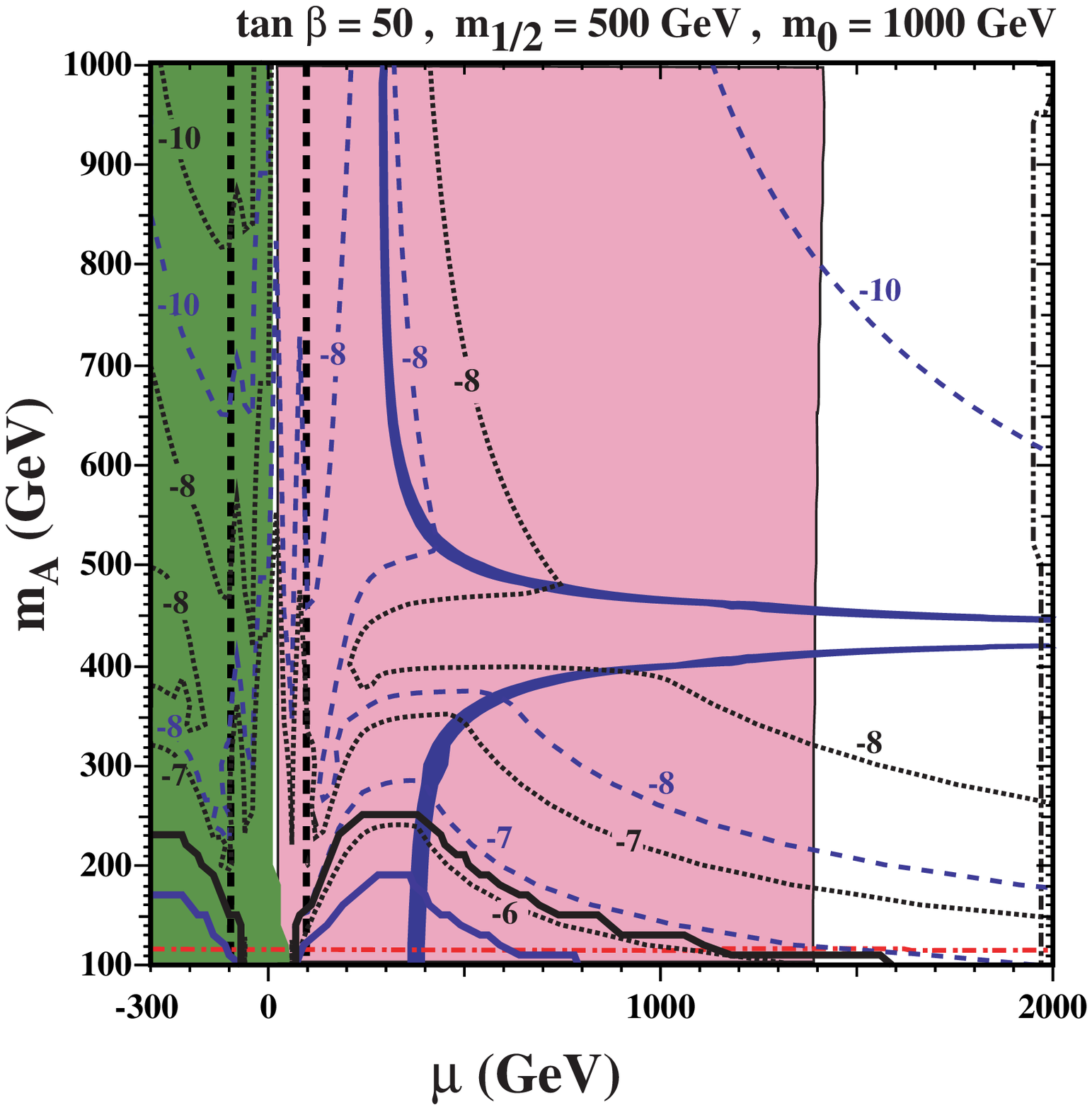,height=8cm}}
\end{center}
\caption{\label{fig:CDMS2}
{\it Contours of the spin-independent elastic cross section in the
$(m_A, \mu)$ planes for $(m_{1/2}, m_0) = (500, 1000)$~GeV and $\tan \beta
=$ (a) 10, (b) 20, (c) 35 and (d)
50, for $\Sigma = 45$~MeV (dashed blue lines) and $\Sigma = 64$~MeV 
(black dotted lines), labelled by their exponents in units of picobarns. 
The 
regions excluded by CDMS~II~\protect\cite{CDMS2} lie below the solid black 
lines.}}
\end{figure}

As usual, the dark, green shaded regions are excluded by $b \to s \gamma$,
the light, pink shaded regions are those preferred by $g_\mu - 2$, the 
solid dark
blue strips are those where $\Omega_\chi$ falls within the range preferred
by WMAP, the red dash-dotted line is the Higgs mass constraint and the
black dashed line is that imposed by the chargino mass. The outward
bulges in the WMAP strips are caused by rapid-annihilation funnels. The
Higgs constraint forbids regions with low $m_A$, which are also excluded
by the GUT Higgs stability constraint for $\tan \beta = 10$, as seen in
Fig.~\ref{fig:CDMS2}(a), but not necessarily for larger values of $\tan
\beta$, as seen in the other panels of Fig.~\ref{fig:CDMS2}. The chargino
constraint removes regions with small $|\mu|$.

Contours of the spin-independent elastic cross section are also plotted in
the $(m_A, \mu)$ planes for various values of $\tan \beta $ in
Fig.~\ref{fig:CDMS2}, labelled by the exponents in units of picobarns 
(blue dashed curves for $\Sigma = 45$~MeV, black dotted curves for $\Sigma 
= 64$~MeV). We see that the largest values of the
spin-independent elastic scattering cross section occur when $\mu$ and
$m_A$ are relatively small. Also displayed in Fig.~\ref{fig:CDMS2} are the
regions excluded by the CDMS~II upper limit (solid black line), including
also the factor $f_\chi < 1$ where appropriate for models with
$\Omega_\chi < \Omega_{CDM}$. In panel (a) for $\tan \beta = 10$, the
regions excluded by CDMS~II were already excluded by the GUT Higgs
stability and $b \to s \gamma$ constraints. However, in the other panels
we see that there are regions at low $\mu$ and $m_A$ that were
allowed by the other constraints but are now excluded by CDMS~II. These 
regions become progressively more extensive as $\tan \beta$ increases.

These regions are reflected in Fig.~\ref{fig:X}(a), which displays in the
$(\mu, m_A)$ plane the NUHM points from Fig.~\ref{fig:NUHM}(c, d) that are
excluded by the CDMS~II result if one assumes $\Sigma = 64$~MeV (dark, red
$\times$ signs) or $\Sigma = 45$~MeV (lighter, blue squares). As
expected, they cluster at small values of $\mu$ and
$m_A$~\footnote{Analogous high-cross-section points for $\mu < 0$ are
excluded by the $b \to s \gamma$ constraint, as seen in 
Fig.~\ref{fig:CDMS2}.}. Their values of $\mu$ and $m_A$ are generally 
smaller than those of the benchmark points~\cite{Bench2}, which are all 
compatible with CDMS~II, as we saw in Fig.~\ref{fig:sigma}. For 
comparison, only
benchmark point B ($m_A \sim 370$ GeV) has a pseudoscalar mass less than
400 GeV and all but points B, I ($m_A \sim 450$ GeV), and L ($m_A \sim
490$ GeV) have pseudoscalar masses in excess of 500 GeV.  Similarly, with
the exception of the focus points (E and F), typical values $\mu$ are
relatively large. Point B has $\mu \sim 330$ GeV, point I has $\mu \simeq
440$ GeV, and G has $\mu \simeq 470$ GeV, whereas all other points have
$\mu$ in excess of 500 GeV. Fig.~\ref{fig:X}(b)  is the corresponding plot
for the excluded LEEST points from Fig.~\ref{fig:LEEST}(c, d). This
exhibits very similar features, confirming the importance of these
variables also in the LEEST scenario. In contrast, the ratios $m_{\tilde
q} / m_{\tilde \ell}$ for the excluded LEEST points do not exhibit any
clustering at low values.

\begin{figure}
\begin{center}
\mbox{\epsfig{file=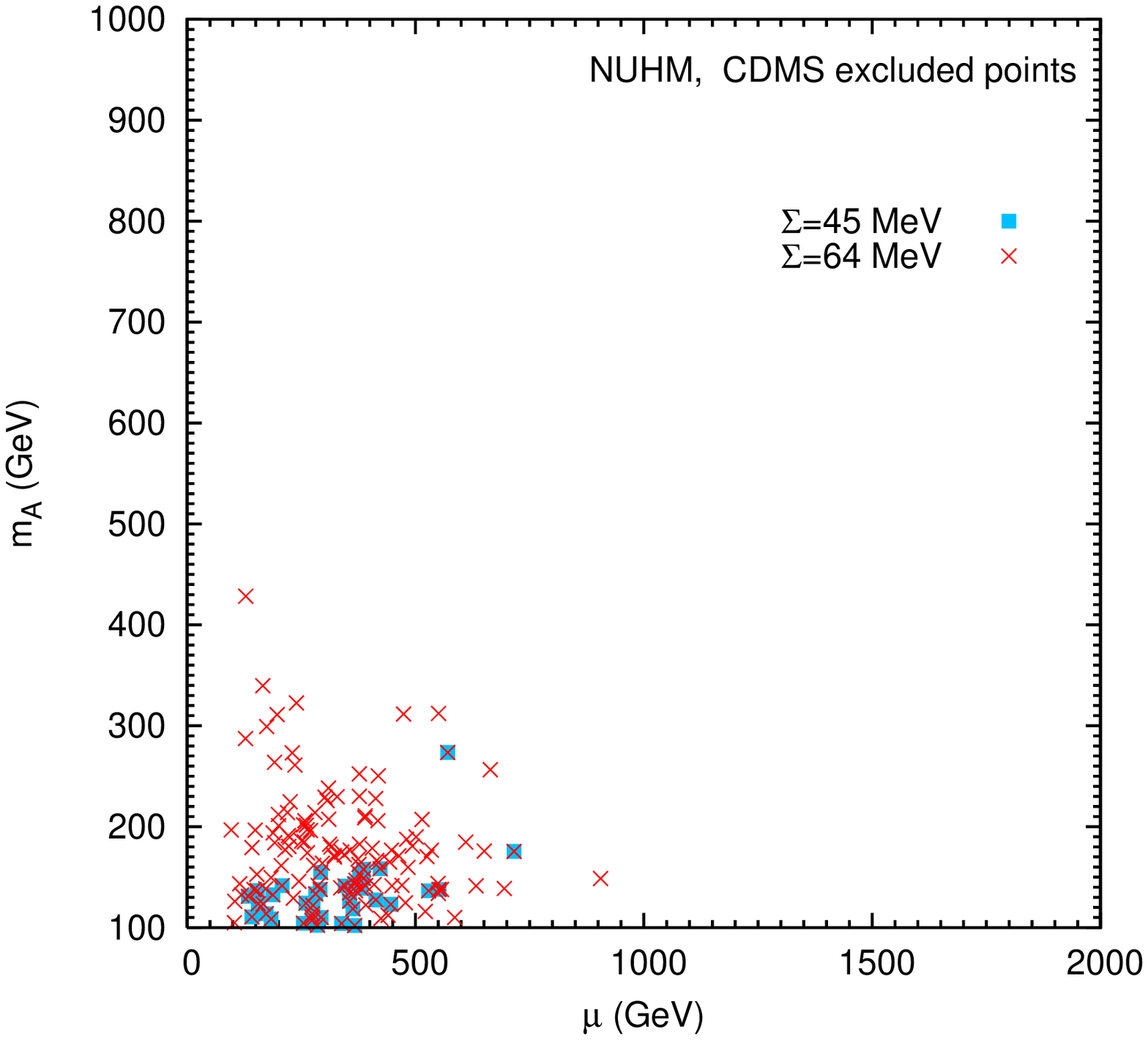,height=9cm}}
\mbox{\epsfig{file=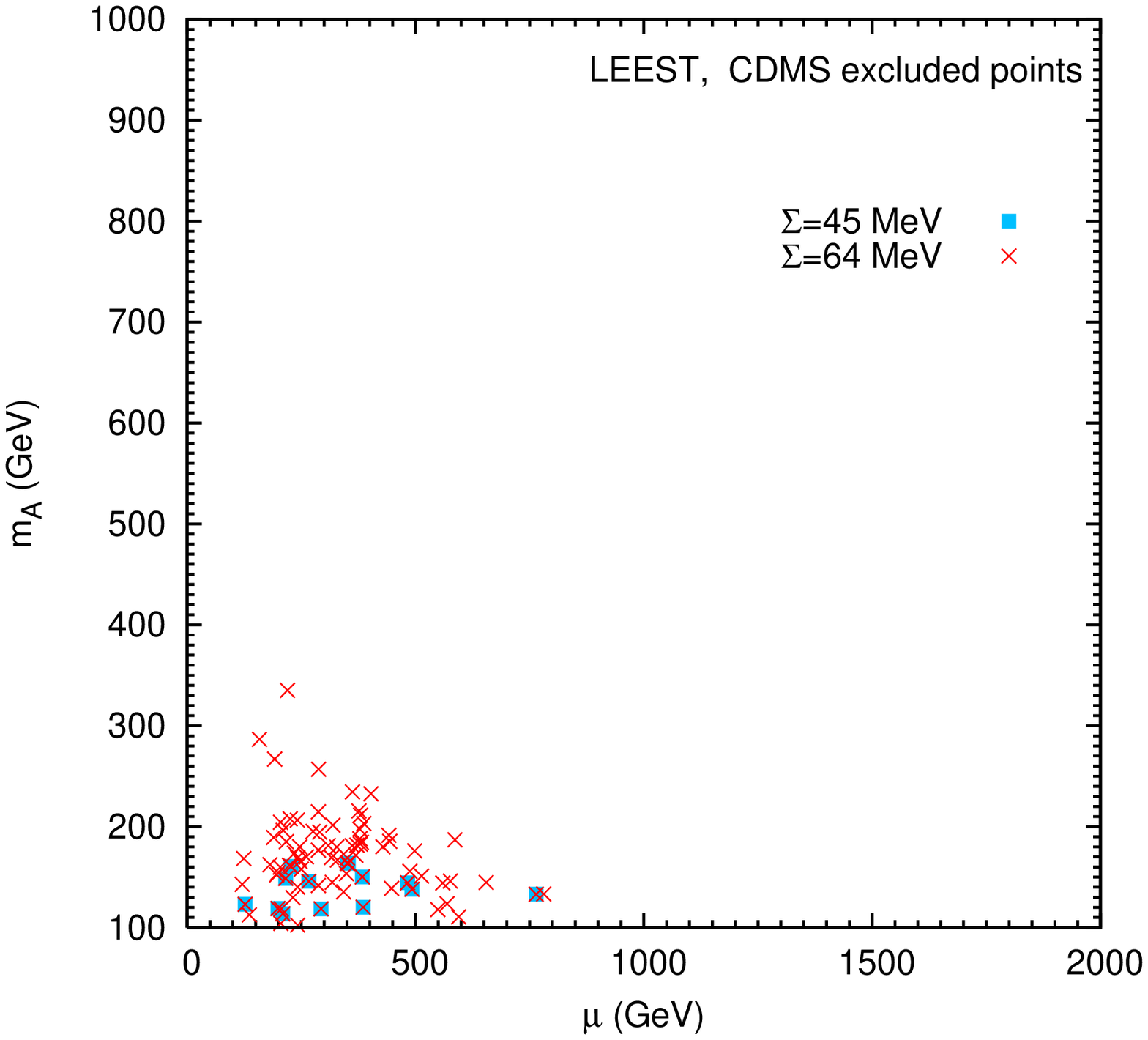,height=9cm}}
\end{center}
\caption{\label{fig:X}
{\it Scatter plots in the $(\mu, m_A)$ plane of points from (a)
Fig.~\ref{fig:NUHM}(c, d) and (b) Fig.~\ref{fig:LEEST}(c, d) that are 
excluded  
by the CDMS~II constraint for  $\Sigma = 64$~MeV (dark, red
$\times$ signs) or $\Sigma = 45$~MeV (lighter, blue squares), 
exhibiting 
similar clustering at low values of
$\mu$ and $m_A$.}}
\end{figure}

\section{Conclusions and Prospects}

In this paper, we have made a new comparison between theoretical
predictions of the spin-independent cross section for the elastic
scattering of supersymmetric dark matter and the improved experimental
upper limit recently provided by CDMS~II~\cite{CDMS2}. In making this
comparison, we have contrasted the theoretical predictions made with
different estimates of the $\pi$-nucleon $\Sigma$ term. Larger values may
be supported by recent reports of exotic baryons, but these do not
increase greatly the ranges of theoretical models excluded by CDMS~II. We
have also incorporated in our analysis the new central value of $m_t$,
which enters indirectly into constraints on the supersymmetric parameter
space and into relic-density calculations.

Some supersymmetric models with non-universal Higgs masses (NUHM) are now
excluded by the CDMS~II upper limit, as are some models which also
incorporate non-universal squark and slepton masses (LEEST). These are
mainly models with the smaller values of $|\mu|$ and/or $m_A$ that become
allowed when the universality conditions are relaxed for the Higgs masses.

On the other hand, only very small parts of the CMSSM parameter space are
yet excluded. Specifically, the cross sections we find in the
supersymmetric benchmark scenarios of~\cite{Bench2} all lie considerably
below the CDMS~II sensitivity, as do all points allowed at the 68\% or
even 90\% confidence level by a recent likelihood~\cite{EHOW3} analysis of
the CMSSM parameter space incorporating information on $m_W, \ \sin^2
\theta_{eff}$ and $g_\mu - 2$.

An improvement over the present CDMS~II sensitivity by about an order of 
magnitude would begin to challenge the preferred region of CMSSM parameter 
space, but an improvement by about four orders of magnitude would be 
required to cover it completely. We conclude that direct searches for 
supersymmetric dark matter are just beginning to reach interesting 
sensitivities, but that considerable improvement will be needed to exclude 
(or hopefully discover) supersymmetric dark matter.

\vspace*{1cm} 
\noindent{ {\bf Acknowledgments} } \\ 
\noindent 
The work of K.A.O., and V.C.S. was supported in part
by DOE grant DE--FG02--94ER--40823. We would like to thank Rick Gaitskell for
information regarding CDMS~II data. Y.S. would like to thank M. Voloshin 
for helpful discussions and the Perimeter Institute for its hospitality.  

\end{document}